\documentstyle[12pt]{article}
\begin{document}
\begin{flushright}
hep-th/0306182\\
AS-ICTP-2003
\end{flushright}
\vskip 2.5cm
\begin{center}
{\bf \Large { Superfield Approach to (Non-)local Symmetries \\
for One-Form Abelian Gauge Theory}}

\vskip 3cm

{\bf R.P.Malik}
\footnote{ E-mail address:  malik@boson.bose.res.in  }
\footnote{ Permanent Address: S. N. Bose National Centre for Basic
Sciences,
Block-JD, Sector-III, Salt Lake, Kolkata--700 098, West Bengal, India.}\\
{\it The Abdus Salam International Centre for Theoretical Physics, }\\
{\it Strada Costiera 11, 34014 Trieste, Italy}\\

\vskip 3cm

\end{center}

\noindent
{\bf Abstract}:
We exploit the geometrical
superfield formalism to derive the local, covariant and continuous
Becchi-Rouet-Stora-Tyutin (BRST) symmetry transformations and the
non-local,
non-covariant and continuous dual-BRST symmetry transformations for the
free Abelian
one-form gauge theory in four $(3 + 1)$-dimensions (4D) of spacetime. Our
discussion is
carried out in the framework of BRST invariant Lagrangian density for the
above
4D theory in the Feynman gauge. The geometrical origin and interpretation
for the (dual-)BRST charges (and the transformations they generate) are
provided in the
language of translations of some superfields along the Grassmannian
directions of
the six ($ 4 + 2)$-dimensional supermanifold parametrized by the four
spacetime
and two Grassmannian  variables. \\

\vskip 0.5cm





\baselineskip=16pt


\newpage

\noindent
{\bf 1 Introduction}\\

\noindent
In the realm of modern developments in
theoretical high energy physics, the symmetry transformations
(and corresponding generators) have played a very important role. In
particular,
the local, covariant and continuous {\it gauge} symmetry transformations
have been
found to dictate the theoretical description of three (out of four)
fundamental interactions of nature. The quantum electrodynamics (QED) is
one of
the most extensively studied {\it gauge theories} where the
experimental tests and theoretical predictions have matched each-other
with an unprecedented degree of 
accuracy in the history of science. One of most
elegant
ways to covariantly quantize such gauge theories (eg QED) is the BRST
formalism
where the ``quantum''
gauge (ie BRST) invariance and unitarity are respected together at any
arbitrary
order of perturbation theory. In this formalism, the local gauge invariant
singular Lagrangian density is extended to include the gauge-fixing and
Faddeev-Popov ghost terms. The ensuing Lagrangian density turns out to be
endowed with a local, covariant, continuous and nilpotent symmetry
transformation which is popularly known as the BRST (or ``quantum'' gauge)
symmetry
transformation [1,2]. Under this transformation, {\it the kinetic energy
term} corresponding to the gauge field of the
Lagrangian density remains invariant (as is the case, even with the usual
local gauge
symmetry transformation). In the recent past, the (anti-)BRST invariant
Lagrangian density
for the one-form (non-)Abelian
gauge theories in 4D has been shown to possess a new nilpotent,
continuous, non-local and non-covariant BRST type transformations under
which
{\it the gauge-fixing term} for the gauge field
remains invariant [3-6]. We christen this
latter symmetry
transformation as the dual(co)-BRST symmetry transformation.
This is because of the fact that there exists a deep connection between the
kinetic
energy term and the gauge-fixing term of the
(anti-)BRST invariant Lagrangian density on the one hand and  the
de Rham cohomological operators of the differential geometry on the other
hand.
For instance, the two-form $F = d A$ defines the curvature term
$F_{\mu\nu}$ (ie $ F = \frac{1}{2}\; (dx^\mu \wedge dx^\nu) \; F_{\mu\nu}$)
from which the kinetic energy term is constructed
and the zero-form $\delta A = - * d * A = (\partial \cdot A)$
implies the existence of $(\partial \cdot A)$ which is responsible for the
construction
of the gauge-fixing term. Here $\delta = - * d *$
(with $\delta^2 =0 )$ and $d = dx^\mu \partial_\mu$
(with $d^2  = 0)$  are the (co-)exterior derivatives and $*$ is the Hodge
duality
operation of the differential geometry (see, eg, [7-11]). Thus,
the kinetic energy term and the gauge-fixing term owe their origin
to the application of $d$ and $\delta$ on the one form $ A = dx^\mu A_\mu$
in a
subtle way. Together with the Laplacian
operator $\Delta = d \delta + \delta d$, the (co-)exterior derivatives
$(\delta)d$  form a set ($ d, \delta, \Delta)$
which is popularly known as the set of de Rham cohomological
operators. These operators obey an algebra: $ d^2 = 0, \delta^2 = 0,
\Delta = (d + \delta)^2 = \{ d, \delta \}, [\Delta, d ] = 0, [\Delta,
\delta] = 0$
showing that $\Delta$ is the Casimir operator
(see, eg, [7,8] for details). The operation of $\Delta$ on the 1-form $A$
(ie $\Delta A = dx^\mu \Box A_\mu$) leads to the derivation of the equation
of motion $\Box A_\mu = 0$ for the gauge-fixed Lagrangian density if we
demand the
validity of Laplace equation $ \Delta A = 0$ for this 1-form gauge theory.

One of the most interesting geometrical approaches to gain an insight into
the
BRST formalism is the superfield formalism [12-17]. In this approach, the
super
exterior derivative $\tilde d$ and the Maurer-Cartan equation are exploited
together
in the so-called horizontality condition
\footnote{ This condition is referred to as the ``soul flatness'' condition
by Nakanishi and Ojima [18].} where the curvature ($(p + 1)$-form) tensor
for
the $p$-form ($ p = 1, 2, 3...$) gauge theory
is restricted to be flat along the Grassmannian
directions of the $(D + 2)$ dimensional supermanifold that is parametrized
by
$D$-number of commuting spacetime variables $x^\mu$ and two
anticommuting
(ie $ \theta^2 = 0, \bar \theta^2 = 0, \theta \bar \theta + \bar \theta
\theta = 0$)
Grassmann variables $\theta$ and $\bar \theta$. This technique leads to the
geometrical interpretation of the conserved and nilpotent ($Q_{(a)b}^2 =
0$)
(anti-)BRST charges $(Q_{(a)b})$ as the translation generators
($\partial/\partial\theta, \partial/\partial \bar\theta$) along the
Grassmannian directions of the supermanifold. Recently, in a set of papers
[19-24], all the super de Rham cohomological operators ($\tilde d, \tilde
\delta,
\tilde \Delta$) have been exploited in the generalized versions of the
horizontality
condition for the 2D free Abelian and self-interacting non-Abelian gauge
theories
on a four $(2 + 2)$-dimensional supermanifold. In this endeavour, the
geometrical
interpretation for (i) the (anti-)BRST charges and corresponding
transformations
\footnote{ We follow here the conventions and notations used by Weinberg
[25]. To
be precise, in their totality, the nilpotent ($\delta_{(A)B}^2 = 0$)
(anti-)BRST transformations $\delta_{(A)B}$ are product of an
anticommuting ($ \eta C + C \eta = 0, \eta \bar C + \bar C \eta = 0$)
spacetime
independent parameter $\eta$ and $s_{(a)b}$ (ie $\delta_{(A)B} = \eta
s_{(a)b}$)
with $s_{(a)b}^2 = 0$.} $s_{(a)b}$, (ii) the (anti-)co-BRST charges and the
transformations $s_{(a)d}$ they generate (iii) a
bosonic charge (ie the anti-commutator of (anti-)BRST and (anti-)co-BRST
charges) and corresponding symmetry transformations $s_{w}$, (iv) the
nilpotency ($Q_{(a)b}^2 = Q_{(a)d}^2 = 0$) of the (anti-)BRST ($Q_{(a)b}$)
and
(anti-)co-BRST ($Q_{(a)d}$) charges, and (v) topological properties of the
above 2D one-form gauge theories, etc, have been provided in the framework
of
superfield formulation. It is interesting to point out that, for the first
time,
the Lagrangian density and symmetric
energy-momentum tensor for the above {\it topological field theories}
have been shown to correspond
to the translation of some composite superfields along the Grassmannian
directions
of the $(2 + 2)$-dimensional supermanifold.

As pointed out earlier, the co-BRST symmetry transformations
are non-local, non-covariant,
continuous and nilpotent [3-6]. Such kind of transformations
(and corresponding non-local generators) have not yet been
discussed in the purview of the geometrical superfield approach to BRST
formalism.
The purpose of the present paper is to provide geometrical origin and
interpretation for the conserved and nilpotent
(co-)BRST charges ($Q_{(d)b}$) (and the transformations they generate) in
the framework of superfield formulation applied to the 4D free 
as well as interacting Abelian
(one-form) gauge theory defined on a six $(4 + 2)$-dimensional
supermanifold.
In particular, it is a challenging endeavour to provide geometrical origin
for the non-local, conserved and nilpotent (anti-)co-BRST charges 
\footnote{ It will be noted that such (anti-)co-BRST charges exist
for the free as well as interacting 4D Abelian one-form gauge theories
and they carry the same expression (cf eqn (2.7) below) for both these cases.}
in the
framework
of superfield formalism as, to the best of our knowledge, such kind of
charges have not yet been discussed in its framework. In the present work,
we
exploit the super (co-)exterior derivatives ($\tilde \delta) \tilde d$
in the (dual-)horizontality conditions on the $(4 + 2)$-dimensional
supermanifold
and demonstrate that the off-shell nilpotent (anti-)BRST charges
(and the nilpotent $\tilde s_{(a)b}^2 = 0$ transformations they generate)
correspond to the translations of some
superfields along the $(\theta)\bar\theta$ directions of the supermanifold.
In the similar fashion, we show that the
off-shell nilpotent (anti-)co-BRST charges (and the nilpotent
$\tilde s_{(a)d}^2 = 0$ transformations they generate) {\it too} correspond
to
the translation of some superfields along the $(\theta)\bar\theta$
directions of the $(4 + 2)$-dimensional supermanifold. However, there is a
clear-cut
distinction between them when it comes to the transformations on the
fermionic (anti-)ghost fields. Whereas the
superfield corresponding to the anti-ghost field $\bar C$ becomes  chiral
under the BRST transformation, it is the superfield
corresponding to the ghost field $C$ that turns into chiral
under the co-BRST transformation. Just the reverse happens when we
consider the derivation of anti-BRST and anti-co-BRST transformations in
the framework
of superfield formulation. In fact, the superfields corresponding to the
(anti-)ghost
fields become anti-chiral in the latter case of superfield formulation.
In the computation of the Hodge duality $\star$ operation on the six
dimensional
supermanifold, we have explained all the steps of our calculation because,
to the
best of our knowledge, this operation is not quite well-known in literature
\footnote{Private communication with V. A. Soroka on this topic is
gratefully
acknowledged.}. We have collected some of the nontrivial results of the
$\star$ operation
in the Appendix too. For the
discussion of the geometrical origin of the on-shell nilpotent (anti-)BRST
and
(anti-)co-BRST transformations, we invoke the (anti-)chiral superfields and
establish
that the on-shell nilpotent charges correspond to the translation of some
variety of
the (anti-)chiral superfields along a specific Grassmannian
direction of the above supermanifold. In fact,
the process of translation of the (anti-)chiral superfields along the
Grassmannian
directions leads to the derivation of internal on-shell nilpotent
symmetries $s_{(a)b}$ and
$s_{(a)d}$ on the basic fields of the Lagrangian density of the 4D free
Abelian
gauge theory. The nilpotency of the on-shell as well as off-shell versions
of
these charges is captured in the property
$(\partial/ \partial \theta)^2 = 0, (\partial/\partial \bar \theta)^2
= 0$ of the translation generators $(\partial /\partial \theta)$ and
$(\partial /\partial \bar \theta)$ along the Grassmannian directions of the
supermanifold.

The outline of our present paper is as follows. In section 2, we briefly
recapitulate
the bare essentials of the (anti-)BRST and (anti-)co-BRST symmetries in the
Lagrangian
formulations to set up the notations and conventions. Section 3 is devoted
to the derivation of the off-shell nilpotent (anti-)BRST and (anti-)co-BRST
symmetries in the framework of superfield formulation. The on-shell
nilpotent (co-)BRST
symmetries are derived by  invoking the chiral superfields in section 4.
Section 5 deals
with the derivation of the on-shell nilpotent anti-BRST and anti-co-BRST
symmetries by
exploiting the anti-chiral superfields. In section 6, the on-shell
nilpotent
(anti-)BRST and (anti-)co-BRST symmetries are deduced together by utilizing
the general
superfield expansions. Finally, in section 7, we make some concluding
remarks and point
out a few future directions that can be pursued later.\\

\noindent
{\bf 2 Preliminary: (co-)BRST symmetries} \\

\noindent
We discuss here the on-shell as well as off-shell nilpotent (anti-)BRST
and (anti-)co-BRST symmetries in the Lagrangian formalism.
To this end in mind, first we begin with the following BRST invariant
Lagrangian density for the four $(3 + 1)$-dimensional (4D) {\it interacting}
Abelian gauge theory
\footnote{ The free 4D Abelian (1-form) gauge theory is the special case
of an interacting theory. 
We adopt here the conventions and notations 
such that the 4D flat Minkowski
manifold is endowed with a metric: $\eta_{\mu\nu} =$ diag $ (+1, -1, -1,
-1)$ and totally
antisymmetric Levi-Civita tensor satisfies $\varepsilon_{\mu\nu\lambda\xi}
\varepsilon^{\mu\nu\lambda\xi} = - 4!, \varepsilon_{\mu\nu\lambda\xi}
\varepsilon^{\mu\nu\lambda\rho} = - 3! \delta^{\rho}_\xi$ etc with the
choice
$\varepsilon_{0123} = + 1, \varepsilon_{0ijk} = \varepsilon_{ijk} = -
\varepsilon^{0ijk}$. Here the Greek indices correspond to spacetime
directions
of the 4D manifold and Latin indices stand for the space directions only.
The 3-vectors
on the manifold are occasionally denoted by the bold faced letters
(eg ${\bf E}$, ${\bf B}$, ${\bf b^{(1)}, b^{(2)}}$).} 
in the Feynman gauge (see, eg, [25-28])
$$
\begin{array}{lcl}
{\cal L}_{b} &=& - \frac{1}{4}\; F^{\mu\nu} F_{\mu\nu}
- \frac{1}{2 }\; (\partial \cdot A)^2 
+ \bar \psi (i \gamma^\mu D_\mu - m) \psi
- i \;\partial_{\mu} \bar C\; \partial^\mu C \nonumber\\
&\equiv& \frac{1}{2}\; ({\bf E^2} - {\bf B^2})
- \frac{1}{2} \; (\partial \cdot  A)^2 
+ \bar \psi (i \gamma^\mu D_\mu - m) \psi
- i \;\partial_{\mu} \bar C \; \partial^\mu C
\end{array} \eqno(2.1)
$$
where $D_\mu = \partial_\mu + i e A_\mu$ is the covariant derivative,
$F_{\mu\nu} = \partial_\mu A_\nu - \partial_\nu A_\mu$ is the
curvature
(field-strength) tensor constructed from the vector potential $A_\mu$ (with
the components $F_{0i} = E_i \equiv {\bf E}, F_{ij} = \varepsilon_{ijk}
B_k,
B_i = \frac{1}{2} \varepsilon_{ijk} F_{jk} \equiv {\bf B}$),
anticommuting
($C^2 = \bar C^2 = 0, C \bar C + \bar C C = 0,
C \psi + \psi C = 0$, etc) (anti-)ghost fields are
required
in the theory to maintain the unitarity and ``quantum'' gauge invariance
together
\footnote{ The true strength of the BRST formalism and its (anti-)ghost
fields turn
up in their full glory for the proof of unitarity in the context of
interacting
non-Abelian gauge theory where there is a gauge
invariant interaction between the quark and gluon
fields (see, eg, [29] for details).} at any arbitrary order of perturbation
theory and $(\bar\psi)\psi$ are the Dirac fields with charge $e$ and mass $m$
(see, eg, [29] for details). As pointed out
earlier in the introduction, we have the gauge-fixing term, the vector
potential
and the curvature term as the zero-form ($\delta A = - * d * A
= (\partial \cdot A)$), 1-form ($ A = dx^\mu A_\mu $) and 2-form
($F = d A = \frac{1}{2} (dx^\mu \wedge dx^\nu)\; F_{\mu\nu}$)
in our present 4D free (one-form) Abelian gauge theory. The gauge-fixing
term
and the curvature 2-form are  constructed
by the application of de Rham cohomological operators $\delta$ and  $d$ on
the 1-form $A = d x^\mu A_\mu$. It is straightforward  to check that
under the following on-shell ($ \Box C = \Box \bar C = 0$)
nilpotent $s_{(a)b}^2 = 0$ (anti-)BRST transformations (with
$ s_b s_{ab} + s_{ab} s_b = 0$) (see, eg, [25-28] for details)
$$
\begin{array}{lcl}
s_{b} A_{\mu} &=& \partial_{\mu} C \quad s_{b} C = 0 \quad
s_{b} \bar C = - i\;(\partial \cdot A) \quad 
s_{b} \psi = -i e C \psi \quad s_{b}\bar \psi = - i e \bar \psi C\nonumber\\
s_{ab} A_{\mu} &=& \partial_{\mu} \bar C\; \quad
s_{ab} \bar C = 0 \quad
s_{ab} C = + i (\partial \cdot A) \quad
s_{ab} \psi = -i e \bar C \psi \quad 
s_{ab}\bar \psi = - i e \bar \psi \bar C
\end{array}\eqno(2.2)
$$
the kinetic energy term
of the Lagrangian density remains invariant. More precisely, the
curvature tensor $F_{\mu\nu}$ itself
remains unchanged under the above transformations. In other words, the
classical
electric field ${\bf E}$ and magnetic field ${\bf B}$ are left intact
under the above nilpotent (anti-)BRST transformations. On the contrary,
under the
following on-shell ($ \Box C = \Box \bar C = 0$) nilpotent ($s_{(a)d}^2 =
0$)
(anti-)co-BRST $s_{(a)d}$ transformations (with $s_d s_{ad} + s_{ad} s_d =
0$)
(see, eg, [3] for details)
$$
\begin{array}{lcl}
s_{d} A_0 &=& i \bar C \quad \;s_{d} A_i = i
{\displaystyle \frac{\partial_0 \partial_i}{\nabla^2}} \bar C \quad\;
s_{d} \bar C = 0 \qquad
s_{d} \psi = \Bigl ({\displaystyle \frac{e}{\nabla^2}} \partial_{0} \bar C
\Bigr ) \;\psi \nonumber\\
 \;s_{d} C &=& - A_0 +
{\displaystyle \frac{\partial_0 \partial_i}{\nabla^2}}\; A_i 
+ {\displaystyle \frac{e}{\nabla^2}} \bar \psi \gamma_{0} \psi \equiv
{\displaystyle \frac{\partial_i E_i + e J_{0}}{\nabla^2}}, \qquad
s_{d} \bar \psi =  \bar \psi \; \Bigl (
{\displaystyle \frac {e}{\nabla^2}}
\partial_{0} \bar C \Bigr ) \nonumber\\
s_{ad} A_0 &=& i C \quad \;s_{ad} A_i = i
{\displaystyle \frac{\partial_0 \partial_i}{\nabla^2}}  C \quad \;
s_{ad} C = 0 \qquad
s_{ad} \psi = \Bigl ({\displaystyle \frac{e}{\nabla^2}} \partial_{0} C
\Bigr ) \;\psi \nonumber\\
 \; s_{ad} \bar C &=&  A_0 -
{\displaystyle \frac{\partial_0 \partial_i}{\nabla^2}}\; A_i 
- {\displaystyle \frac{e}{\nabla^2}} \bar\psi \gamma_{0} \psi \equiv
- {\displaystyle \frac{\partial_i E_i + e J_{0}}{\nabla^2}} \quad
s_{ad} \bar \psi =  \bar \psi \; \Bigl (
{\displaystyle \frac {e}{\nabla^2}}
\partial_{0}  C \Bigr ) 
\end{array}\eqno(2.3)
$$
it is the gauge-fixing term that remains invariant. More precisely, the
term
$(\partial \cdot A)$ itself
remains unchanged under the above transformation. The salient features, at
this stage,
are (i) the above Lagrangian density remains invariant (modulo a total
derivative)
under the (anti-)BRST as well as the (anti-)co-BRST transformations. (ii)
The (anti-)BRST
transformations leave the 2-form $F = d A$ invariant. (iii) The
(anti-)co-BRST
transformations keep the zero-form $ (\partial \cdot A) = \delta A$
unaltered. (iv)
The 2-form $F= d A$ and the zero-form $\delta A = (\partial \cdot A)$ are,
in some
sense, ``Hodge dual'' to each-other because $\delta = - * d *$ and $d$ are
Hodge dual to each-other. (v) The magnetic field ${\bf B}$ remains
invariant
(ie $s_{(a)b} B_i = s_{(a)d} B_i = 0$) under
all the nilpotent (anti-)BRST and (anti-)co-BRST transformations. (vi) It
is
obvious that a bosonic symmetry $s_w$ can be obtained from the
anticommutator
$(s_{w} = \{ s_{b}, s_{d} \} = \{ s_{ab}, s_{ad} \}$) of the
(anti-)BRST $s_{(a)b}$ and (anti-)co-BRST $s_{(a)d}$ symmetries. However,
we shall
not discuss here about this symmetry as it is not essential for
our present work. An elementary discussion
on it can be found in [3] (vii) The (anti-)BRST transformations are local,
continuous,
covariant and nilpotent but the (anti-)co-BRST transformations are
non-local,
continuous, non-covariant and nilpotent. (viii) The off-shell nilpotent
version of the
above nilpotent symmetries has not been discussed {\it together}
in [3-6]. We obtain the off-shell nilpotent version of the above symmetries
by
invoking a couple of 3-vector auxiliary fields ${\bf b^{(1)}}$ and ${\bf
b^{(2)}}$ and a
scalar auxiliary field $b_3$. They play an important role in linearizing
the Lagrangian
density (2.1) and, in the  process, change it to the following form
$$
\begin{array}{lcl}
{\cal L}_{B} &=& b^{(1)}_i E_i - {\displaystyle \frac{1}{2}}\; {\bf
(b^{(1)})^2}
- b^{(2)}_i B_i + {\displaystyle \frac{1}{2}}\;{\bf (b^{(2)})^2}
+ b_{3} (\partial \cdot A)  + {\displaystyle\frac{1}{2}}\;(b_{3})^2
\nonumber\\
&+& \bar \psi (i \gamma^\mu D_\mu - m) \psi
- i \;\partial_{\mu} \bar C \; \partial^\mu C.
\end{array} \eqno(2.4)
$$
It is straightforward to check that ${\bf b^{(1)} = E}$,  ${\bf b^{(2)} =
B}$
and $b_{3} = - (\partial \cdot A)$. The off-shell nilpotent version of the
(anti-)BRST transformations (2.2) is
$$
\begin{array}{lcl}
\tilde s_{b} A_{\mu} &=& \partial_{\mu} C \quad \tilde s_{b} C = 0\;
\quad
\tilde s_{b} \psi = -i e C \psi \quad 
\tilde s_{b} \bar C = i\;b_{3} \qquad \tilde s_{b} b_3 = 0 \nonumber\\
\tilde s_{b} {\bf E} &=& 0 \qquad \;\tilde s_{b} {\bf B} = 0 \qquad\;
\tilde s_{b} {\bf b^{(1)}} = 0 \qquad \; \tilde s_{b} {\bf b^{(2)}} = 0
\quad \tilde s_{b}\bar \psi = - i e \bar \psi C 
\nonumber\\
\tilde s_{ab} A_{\mu} &=& \partial_{\mu} \bar C \quad 
\tilde s_{ab} \bar C = 0 \quad 
\tilde s_{ab} \psi = -i e \bar C \psi \quad 
\tilde s_{ab} C = - i\;b_3 \; \qquad \; \tilde s_{ab} b_3 = 0 \nonumber\\
\tilde s_{ab} {\bf E} &=& 0 \;\qquad \;\tilde s_{ab} {\bf B} = 0 \;\qquad
\tilde s_{ab}\bar \psi = - i e \bar \psi \bar C
\quad
\tilde s_{ab} {\bf b^{(1)}} = 0 \;\qquad \; \tilde s_{ab} {\bf b^{(2)}} = 0
\end{array}\eqno(2.5)
$$
and that of the (anti-)co-BRST symmetry transformations in (2.3), is
$$
\begin{array}{lcl}
\tilde  s_{d} A_0 &=& i \bar C \;\qquad \;\tilde s_{d} A_i = i
{\displaystyle \frac{\partial_0 \partial_i}{\nabla^2}} \bar C \;\qquad \;
\tilde s_{d} \bar C = 0 \nonumber\\
\tilde s_{d} C &=&
{\displaystyle \frac{\partial_i b^{(1)}_i + e J_{0}}{\nabla^2}} \qquad
\tilde s_{d} \psi = \Bigl ( {\displaystyle \frac{e}{\nabla^2}}
\partial_{0} \bar C) \; \psi  \quad
\tilde s_{d} \bar \psi =  \bar \psi \; \Bigl (
{\displaystyle \frac {e}{\nabla^2}}
\partial_{0} \bar C \Bigr ) \nonumber\\
\tilde s_{d} {\bf b^{(1)}} &=& 0 \qquad \tilde s_{d} b_3 = 0 \qquad
\tilde s_{d} {\bf b^{(2)}} = 0 \qquad  \tilde s_{d} (\partial \cdot A) = 0
\qquad
\tilde s_{d} {\bf B} = 0 \nonumber\\
\tilde s_{ad} A_0 &=& i C \;\qquad \;\tilde s_{ad} A_i = i
{\displaystyle \frac{\partial_0 \partial_i}{\nabla^2}}  C \;\qquad\;
\tilde s_{ad} C = 0 \nonumber\\
 \;\tilde s_{ad} \bar C &=&  -
{\displaystyle \frac{\partial_i b^{(1)}_i + e J_{0}}{\nabla^2}} \quad
\tilde s_{ad} \psi = \Bigl ({\displaystyle \frac{e}{\nabla^2}} \partial_{0} C
\Bigr ) \;\psi \quad 
\tilde s_{ad} \bar \psi =  \bar \psi \; \Bigl (
{\displaystyle \frac {e}{\nabla^2}}
\partial_{0}  C \Bigr ) 
\nonumber\\
\tilde s_{ad} {\bf b^{(1)}} &=& 0 \qquad \tilde s_{ad} b_3 = 0 \qquad
\tilde s_{ad} {\bf b^{(2)}}
= 0 \qquad  \tilde s_{ad} (\partial \cdot A) = 0 \qquad
\tilde s_{ad} {\bf B} = 0.
\end{array}\eqno(2.6)
$$
In the later sections, we shall see that the auxiliary fields,
present in the Lagrangian density
(2.4) for the derivations of the off-shell nilpotent (anti-)BRST and
(anti-)co-BRST
versions of symmetry transformations, will play important roles.

The (non-)local, conserved and on-shell nilpotent generators for the above
on-shell nilpotent (co-)BRST
transformations can be computed from the Noether conserved current. These, for
the free as well as interacting 4D Abelian gauge theories,
are [3]
$$
\begin{array}{lcl}
Q_{d} &=& {\displaystyle \int}\; d^3 x \;\bigl [\;
{\displaystyle \frac{\partial_0 (\partial \cdot A)} {\nabla^2} } \dot {\bar
C}
- (\partial \cdot A) \bar C \; \bigr ] \nonumber\\
Q_{b} &=& {\displaystyle \int}\; d^3 x \; \bigl [\;
\partial_0 (\partial \cdot A) \;C
- (\partial \cdot A) \dot C \; \bigr ].
\end{array} \eqno(2.7)
$$
From the above expressions, the (non-)local, nilpotent and conserved
charges
corresponding to anti-co-BRST symmetries and anti-BRST symmetries can be
computed
by the substitutions: $ C \rightarrow \pm i \bar C, \bar C \rightarrow \pm
i C$
which turn out to be the discrete symmetry transformations for the ghost
part of
the action. In fact, for the generic field $\Psi (x) = (A_\mu, C, \bar C)
(x)$
of the theory,
the conserved charges $Q_r$  generate the following generic transformations
$$
\begin{array}{lcl}
s_{r} \; \Psi = - i\; \bigl [ \Psi, Q_r \bigr ]_{\pm} \;\;\qquad\;\;
r = b, ab, d, ad, w, g
\end{array} \eqno(2.8)
$$
where $Q_g$ stands for the conserved ghost charge which generates an
infinitesimal
continuous and  global scale transformation for the basic fields of the
theory as:
$ s_g A_\mu = 0, s_{g} C = - \Lambda C, s_g \bar C = + \Lambda \bar C$
where $\Lambda$
is a global parameter. The $(+)-$ signs, as the subscripts on the square
bracket,
imply (anti-)commutators depending on the generic field $\Psi$ being
(fermionic)bosonic
in nature. Thus, we note that (i) there are four (non-)local, continuous,
(non-)covariant and on-shell as well as off-shell
nilpotent (ie, $s_{(a)b}^2 = s_{(a)d}^2 = \tilde s_{(a)b}^2 =
\tilde s_{(a)d}^2 = 0$) symmetries and
a couple of continuous, (non-)local and (non-)covariant bosonic symmetries
$s_w$ and
$s_g$ in the theory, and (ii) the generic transformation in (2.8) is also
valid for
the off-shell nilpotent (anti-)BRST $\tilde s_{(a)b}$ transformations,
(anti-)co-BRST $\tilde s_{(a)d}$ transformations
and the corresponding bosonic $\tilde s_{w} = \{ \tilde s_{(a)b}, \tilde
s_{(a)d} \} $
transformations as well as the scale symmetry transformations $s_{g}$.\\

\noindent
{\bf 3 Off-shell nilpotent symmetries: general superfield approach}\\

\noindent
To derive the off-shell nilpotent (anti-)BRST and (anti-)co-BRST symmetries
{\it together} in the framework of superfield formulation, we resort to the
most
general super expansion for the superfields $B_\mu (x, \theta, \bar\theta),
\Phi (x, \theta, \bar\theta)$ and $\bar \Phi (x,\theta,\bar\theta)$. These
superfields
are the generalization of the local fields $A_\mu (x), C (x)$ and $\bar C
(x)$ of
the 4D Lagrangian density (2.1) to a six $(4 + 2)$-dimensional
supermanifold
which is parametrized by the four bosonic (even) spacetime ($x^\mu, \mu =
0, 1, 2, 3)$ coordinates and two (odd) Grassmannian ($\theta^2
= \bar\theta^2 = 0,
\theta \bar\theta + \bar \theta \theta = 0$) variables. The most general
expansion for the above superfields are
$$
\begin{array}{lcl}
B_{\mu} (x, \theta, \bar \theta) &=& A_{\mu} (x)
+ \theta\; \bar R_{\mu} (x) + \bar \theta\; R_{\mu} (x)
+ i \;\theta \;\bar \theta \;S_{\mu} (x) \nonumber\\
\Phi (x, \theta, \bar \theta) &=& C (x)
+ i\; \theta \; \bar b_3 (x)
+ i \;\bar \theta\; {\cal B} (x)
+ i\; \theta\; \bar \theta \;s (x) \nonumber\\
\bar \Phi (x, \theta, \bar \theta) &=& \bar C (x)
+ i \;\theta\; \bar {\cal B} (x) + i\; \bar \theta \;b_3 (x)
+ i \;\theta \;\bar \theta \;\bar s (x)
\end{array} \eqno(3.1)
$$
where the number of degree of freedom associated with both the sets of the
bosonic fields
$(A_\mu, S_\mu, b_3, \bar b_3, {\cal B}, \bar {\cal B})$ as well as the
fermionic fields $(R_\mu, \bar R_\mu, C, \bar C, s, \bar s)$ are equal. 
It should be noted that the local matter fields $\psi$ and $\bar \psi$
of the Lagrangian density (2.1) have not been generalized to their 
counterparts in the language of superfields. This is due to the fact that,
to the best of our knowledge, it is {\it not} known how to obtain the BRST-type
symmetry transformations on the matter fields in the framework of
superfield formulation. We comment on it in the conclusion (cf section 7)
part of our present paper. The most general form of
the super exterior derivative $\tilde d$ and the one-form super connection
$\tilde A$
$$
\begin{array}{lcl}
\tilde d &=& d Z^M \;\partial_{M} \equiv
d x^\mu \;\partial_\mu \;+\; d \bar \theta\;\partial_{\bar\theta} \;+\;
d \theta \;\partial_{\theta}
\nonumber\\
\tilde  A &=& d Z^M\; \tilde A_{M} \equiv
dx^\mu \;B_\mu (x,\theta,\bar\theta)
+ d \bar \theta\; \Phi (x,\theta,\bar\theta)
+ d \theta\; \bar \Phi (x,\theta,\bar\theta)
\end{array} \eqno(3.2)
$$
(with $Z^M = x^\mu, \theta, \bar\theta$)
lead to the following super curvature 2-form ($\tilde F = \tilde d \tilde
A$)
$$
\begin{array}{lcl}
\tilde d \tilde A &=& \frac{1}{2}\; (d Z^M \wedge d Z
^N)\; F_{MN} \nonumber\\
&\equiv& (dx^\mu \wedge dx^\nu)\; (\partial_\mu B_\nu)\; - \;
(d \theta \wedge d \theta)\;
(\partial_\theta \bar \Phi) \;
- \;(d\theta \wedge d \bar \theta)\; (\partial_\theta \Phi +
\partial_{\bar\theta}
\bar \Phi)\nonumber\\
&-& (d \bar\theta \wedge d \bar\theta) (\partial_{\bar\theta}
\Phi) + (dx^\mu \wedge d\theta) (\partial_\mu \bar \Phi - \partial_\theta
B_\mu)
+ (dx^\mu \wedge d \bar \theta) (\partial_\mu \bar \Phi
- \partial_{\bar\theta} B_\mu).
\end{array} \eqno(3.3)
$$
Now we exploit the horizontality condition ($\tilde d \tilde A = d A$)
which physically implies that there are no superspace contributions to the
physical electric and magnetic fields described by the 2-form
$ F = d A = \frac{1}{2} (dx^\mu \wedge dx^\nu) F_{\mu\nu}$
in the usual 4D spacetime. In other words, all the components of
$F_{MN}$ with Grassmannian variables $\theta$ and/or $\bar \theta$ will
be flat. This results in the following relationships among the auxiliary
fields
of expansion in (3.1) and the basic local fields of the Lagrangian density
(2.1)
(see, eg, [14,20] for details)
$$
\begin{array}{lcl}
&& {\cal B} (x) = \bar {\cal B} (x) = 0 \qquad s (x) =
\bar s(x) = 0 \qquad b_3  (x) + \bar b_3 (x) = 0
\nonumber\\
&& R_\mu (x) = \partial_\mu C (x) \qquad \bar R_\mu (x) = \partial_\mu
\bar C (x) \qquad S_\mu (x) = \partial_\mu b_3 (x).
\end{array} \eqno(3.4)
$$
With the above relationships, the expansion in (3.1) can be re-expressed
in terms of the off-shell nilpotent (anti-)BRST transformations of (2.5) as
$$
\begin{array}{lcl}
B_{\mu} (x, \theta, \bar \theta) &=& A_{\mu} (x)
+ \theta\; (\tilde s_{ab} A_\mu (x)) + \bar \theta\; (\tilde s_{b} A_{\mu}
(x))
+  \;\theta \;\bar \theta\; (\tilde s_{b} \tilde s_{ab} A_{\mu} (x))
\nonumber\\
\Phi (x, \theta, \bar \theta) &=& C (x)
+  \theta \; (\tilde s_{ab} C (x))
+ \;\bar \theta\;(\tilde s_{b} C (x))
+ \; \theta\; \bar \theta \;(\tilde s_{b} \tilde s_{ab} C (x)) \nonumber\\
\bar \Phi (x, \theta, \bar \theta) &=& \bar C (x)
+  \;\theta\; (\tilde s_{ab} \bar C (x)) + \; \bar \theta \;(\tilde s_{b}
\bar C (x))
+  \;\theta \;\bar \theta \; (\tilde s_{b} \tilde s_{ab} \bar C (x)).
\end{array} \eqno(3.5)
$$
The above expansion, in view of the relationships in (2.8) for the
generators of
internal transformations, unequivocally  makes it clear that the local
conserved
and off-shell nilpotent (anti-)BRST charges {\it geometrically} correspond
to the
translation generators $(\partial/\partial \theta) \partial/ \partial
\bar\theta$
along the $(\theta) \bar\theta$ directions of the $(4 + 2)$-dimensional
supermanifold (in the limit $\theta,\bar\theta \rightarrow 0$)
and their off-shell nilpotency is captured in the specific
property of the translation generators
which obey $(\partial / \partial \theta)^2 = 0,
(\partial / \partial \bar\theta)^2 = 0$ (cf section 7).

Now we shall dwell on the derivation of the off-shell nilpotent,
continuous,
non-local and non-covariant (anti-)co-BRST symmetry transformations of
(2.6) in
the framework of superfield formulation. To this end in mind, first of all
we derive the dual version ($\star \tilde A = \star (dZ^M) A_M$)
of the super one-form connection $\tilde A$ defined in (3.2). The resulting
dual super form ($ \star \tilde A$) is, of course, a five-form in
the six ($4 + 2)$-dimensional supermanifold. The explicit expression for
this
$\star$ operation on $\tilde A$ is
$$
\begin{array}{lcl}
&& \star\; \tilde A =
{\displaystyle \frac{1} {3!}}\;\varepsilon^{\mu\nu\lambda\xi}
(dx_\nu \wedge dx_\lambda \wedge dx_\xi \wedge d \theta \wedge
d \bar \theta)\; B_\mu (x, \theta, \bar\theta) \nonumber\\
&& +
{\displaystyle \frac{1} {4!}}\;\varepsilon^{\mu\nu\lambda\xi}
(dx_\mu \wedge dx_\nu \wedge dx_\lambda \wedge dx_\xi \wedge
d \bar \theta)\; \bar \Phi (x, \theta, \bar\theta) \nonumber\\
&& + {\displaystyle \frac{1} {4!}}\;\varepsilon^{\mu\nu\lambda\xi}
(dx_\mu \wedge dx_\nu \wedge dx_\lambda \wedge dx_\xi \wedge d \theta)\;
\Phi (x, \theta, \bar\theta).
\end{array} \eqno(3.6)
$$
In fact,
the above five-form $(\star \tilde A)$ has been computed for the purpose of
the operation of super co-exterior derivative $ \tilde \delta =
- \star\; \tilde d \;\star$ on the super one-form $\tilde A$ where
the $\star$ operation is the Hodge duality operation defined on the
$(4 + 2)$-dimensional supermanifold. The following $\star$ operation on the
super
differentials $ (d Z^M)$ has been taken into account in the above
computation
$$
\begin{array}{lcl}
&&\star\; (dx^\mu) =
{\displaystyle \frac{1} {3!}}\;\varepsilon^{\mu\nu\lambda\xi}\;
(dx_\nu \wedge dx_\lambda \wedge dx_\xi \wedge d \theta \wedge
d \bar \theta) \nonumber\\
&&\star\; (d \theta) =
{\displaystyle \frac{1} {4!}}\;\varepsilon^{\mu\nu\lambda\xi}\;
(dx_\mu \wedge dx_\nu \wedge dx_\lambda \wedge dx_\xi \wedge
d \bar \theta) \nonumber\\
&& \star\; (d \bar\theta) =
{\displaystyle \frac{1} {4!}}\;\varepsilon^{\mu\nu\lambda\xi}\;

(dx_\mu \wedge dx_\nu \wedge dx_\lambda \wedge dx_\xi \wedge d \theta).
\end{array} \eqno(3.7)
$$
It should be noted that (i) in the denominator, the factorials have been
taken corresponding to the 4D spacetime Minkowski manifold because the
Grassmannian
differentials behave in a completely different manner than the spacetime
differentials.
(ii) Even though, the 2-form differentials $d\theta \wedge d \theta$ and
$d \bar\theta \wedge d \bar \theta$ do exist in terms of the Grassmannian
co-ordinates, they have not been taken into account in the $\star$
operation
on the one-form spacetime differential $dx^\mu$. This is because of the
fact that
$\theta$ and $\bar\theta$ directions are the independent directions on the
supermanifold which constitute the dual directions
for the differential $(dx^\mu)$ along with the other
three spacetime directions. The latter spacetime (dual)
directions are taken into account through
the 4D Levi-Civita tensor $\varepsilon^{\mu\nu\lambda\xi}$
defined on the 4D Minkowski manifold. (iii) In the limit $(\theta,
\bar\theta)
\rightarrow 0$, we get back the ordinary Hodge duality
$*$ operation defined on the 4D Minkowski manifold. (iv) We follow the
convention of arranging the spacetime differentials to
the left and Grassmannian differentials to the right in
all the wedge products. Now we apply the super exterior derivative $\tilde
d$
on (3.6) which yields the following
$$
\begin{array}{lcl}
&& \tilde d\;\star\; \tilde A =
{\displaystyle \frac{1} {3!}}\;\varepsilon^{\mu\nu\lambda\xi}\;
(dx_\rho \wedge dx_\nu \wedge dx_\lambda \wedge dx_\xi \wedge d \theta
\wedge
d \bar \theta)\; (\partial^\rho B_\mu) (x,\theta,\bar\theta) \nonumber\\
&&- {\displaystyle \frac{1} {4!}}\;\varepsilon^{\mu\nu\lambda\xi}\;
(dx_\mu \wedge dx_\nu \wedge dx_\lambda \wedge dx_\xi \wedge d\theta \wedge
d \bar \theta)\; (\partial_\theta \bar \Phi) (x,\theta,\bar\theta)
\nonumber\\
&& - {\displaystyle \frac{1} {4!}}\;\varepsilon^{\mu\nu\lambda\xi}\;
(dx_\mu \wedge dx_\nu \wedge dx_\lambda \wedge dx_\xi \wedge d \theta
\wedge d \bar\theta)\; (\partial_\theta \Phi) (x,\theta,\bar\theta)
\nonumber\\
&&- {\displaystyle \frac{1} {4!}}\;\varepsilon^{\mu\nu\lambda\xi}\;
(dx_\mu \wedge dx_\nu \wedge dx_\lambda \wedge dx_\xi \wedge d \bar \theta
\wedge d \bar\theta)\; (\partial_{\bar\theta} \bar \Phi)
(x,\theta,\bar\theta)
\nonumber\\
&& - {\displaystyle \frac{1} {4!}}\;\varepsilon^{\mu\nu\lambda\xi}\;
(dx_\mu \wedge dx_\nu \wedge dx_\lambda \wedge dx_\xi \wedge d \theta
\wedge d \bar\theta)\; (\partial_{\bar \theta} \Phi) (x,\theta,\bar\theta).
\end{array} \eqno(3.8)
$$
A few noteworthy points at this stage are (i) we have dropped all the terms
in the above which possess more than four differentials in terms of the
spacetime co-ordinates and more than two differentials in the Grassmann
variables. (ii) The origin for the existence of the differentials
$ d\theta \wedge d \bar\theta$ in the first term, second term and the last
term
is entirely  different. That is to say, the latter two are similar but
completely
different from the first term in their origin.
Thus, while taking the another $\star$ on (3.8), this
difference will appear. In fact, another  $ \star$ operation
(due to $ \tilde \delta \tilde A = - \star \tilde d \star \tilde A$)
on (3.8) leads to the following
$$
\begin{array}{lcl}
\tilde \delta \tilde A \equiv
- \star\; \tilde d\; \star \tilde A = (\partial \cdot B) -
s^{\theta\bar\theta}\; (\partial_\theta \bar \Phi + \partial_{\bar \theta}
\Phi)
- s^{\theta\theta}\; (\partial_\theta \Phi) - s^{\bar\theta\bar\theta}\;
(\partial_{\bar\theta} \bar \Phi)
\end{array} \eqno(3.9)
$$
where coefficients $s's$ are symmetric
(ie $s^{\theta\bar\theta} = s^{\bar\theta\theta}$ etc) and we have
exploited the
following
$$
\begin{array}{lcl}
&&\star\; (dx_\rho \wedge dx_\nu \wedge dx_\lambda \wedge dx_\xi \wedge
d\theta
\wedge d \bar\theta)\; (\partial^\rho B_\mu)
= \varepsilon_{\rho\nu\lambda\xi} \; (\partial^\rho B_\mu) \nonumber\\
&& \star\; (dx_\mu \wedge dx_\nu \wedge dx_\lambda \wedge dx_\xi \wedge
d\theta
\wedge d \bar\theta)\; (\partial_\theta \bar \Phi)
= \varepsilon_{\mu\nu\lambda\xi}\; s^{\theta\bar\theta}\;
(\partial_\theta \bar \Phi)\nonumber\\
&& \star\; (dx_\mu \wedge dx_\nu \wedge dx_\lambda \wedge dx_\xi \wedge
d\theta
\wedge d \bar\theta)\; (\partial_{\bar\theta} \Phi)
= \varepsilon_{\mu\nu\lambda\xi}\; s^{\theta\bar\theta}\;
(\partial_{\bar\theta} \Phi).
\end{array} \eqno(3.10)
$$
It is evident that the presence of the symmetric $s's$
in the $\star$ operation depends on the origin
of the wedge products $d \theta \wedge d \bar\theta$. This has been done to
account for the property of the duality $\star$ operation which requires
the
validity of $ \star\; (\star \; G) = \pm\; G $ on any generic superfield
$G$
(see, eg, [30] for details on the ordinary $*$ operations). Thus,
the existence of $s's$
keep track of (i) the origin of the Grassmannian differentials, and (ii)
the kind of differentials one should get after a couple
of successive $\star$ operations on any arbitrary differential super forms
(that contain the Grassmann differentials). Some of these $\star$
operations
have been collected in the Appendix. The application of the
dual-horizontality
condition ($ \tilde \delta \tilde A = \delta A$) leads to the following
$$
\begin{array}{lcl}
&&b_3 (x) = \bar b_{3} (x) = s (x) = \bar s (x) = 0 \qquad
{\cal B} (x) + \bar {\cal B} (x) = 0\nonumber\\
&& (\partial \cdot R) (x) = 0 \qquad (\partial \cdot \bar R) (x) = 0 \qquad
(\partial \cdot S) (x) = 0.
\end{array} \eqno(3.11)
$$
It is straightforward to check that the following choices for the free theory
$$
\begin{array}{lcl}
&& R_0 = i \bar C \qquad
R_i = i {\displaystyle \frac{\partial_i \partial_0}
{\nabla^2}}\; \bar C  \qquad \bar R_0 = i C \nonumber\\
&& \bar R_i = i {\displaystyle \frac{\partial_i \partial_0}
{\nabla^2}}\; C \qquad {\cal B} = - i {\displaystyle
\frac{\partial_i b_{i}^{(1)}}{\nabla^2}} \qquad
\bar {\cal B} = + i {\displaystyle
\frac{\partial_i b_{i}^{(1)}}{\nabla^2}}
\end{array} \eqno(3.12)
$$
do satisfy the above conditions emerging from the dual-horizontality
conditions. For the interacting theory, the auxiliary fields can
be chosen as: $ {\cal B}^{(I)} = - i (\partial_i b_{i}^{(1)} 
+ e J_{0})/ \nabla^2, \bar {\cal B}^{(I)} =
+ i (\partial_i b_{i}^{(1)} + e J_{0})/ \nabla^2 $. The expressions
for $R_\mu$ and $\bar R_\mu$ in (3.12) remain intact for the interacting case. 
It is clear that one can {\it not} obtain
the transformations on the matter fields $\psi$ and $\bar \psi$ in the present
form of the superfield formulation.
As far as the determination of $S_\mu(S_\mu^{(I)})$ is concerned, we choose
judiciously  the following expressions for its components in the case of free
and interacting theories
$$
\begin{array}{lcl}
&&S_{0} =  {\displaystyle \frac{\partial_i b_{i}^{(1)}}{\nabla^2}} \qquad\;\;
S_{i} =  {\displaystyle \frac{\partial_i \partial_0}{\nabla^2}}\Bigl (
{\displaystyle \frac{\partial_j b_{j}^{(1)}}{\nabla^2}} \Bigr )\nonumber\\
&& S_{0}^{(I)} =  {\displaystyle \frac{\partial_i b_{i}^{(1)}
+ e J_{0} }{\nabla^2}} \qquad\;\;\;
S_{i}^{(I)} =  {\displaystyle \frac{\partial_i \partial_0}{\nabla^2}}
\Bigl ({\displaystyle \frac{\partial_j b_{j}^{(1)}
+ e J_{0}}{\nabla^2}} \Bigr )
\end{array} \eqno(3.13)
$$
It is worth pointing out that the auxiliary field ${\bf b^{(2)}}$ has not
been
taken into account here because this field and its equivalent (the magnetic
field
${\bf B}$) do not transform under any of the transformations discussed
above.
In terms of the above quantities and the transformations (2.6), we obtain
the
following expansions for the superfields in (3.1)
$$
\begin{array}{lcl}
B_{\mu} (x, \theta, \bar \theta) &=& A_{\mu} (x)
+ \theta\; (\tilde s_{ad} A_\mu (x)) + \bar \theta\; (\tilde s_{d} A_{\mu}
(x))
+  \;\theta \;\bar \theta\; (\tilde s_{d} \tilde s_{ad} A_{\mu} (x))
\nonumber\\
\Phi (x, \theta, \bar \theta) &=& C (x)
+  \theta \; (\tilde s_{ad} C (x))
+ \;\bar \theta\;(\tilde s_{d} C (x))
+ \; \theta\; \bar \theta \;(\tilde s_{d} \tilde s_{ad} C (x)) \nonumber\\
\bar \Phi (x, \theta, \bar \theta) &=& \bar C (x)
+  \;\theta\; (\tilde s_{ad} \bar C (x)) + \; \bar \theta \;(\tilde s_{d}
\bar C (x))
+  \;\theta \;\bar \theta \; (\tilde s_{d} \tilde s_{ad} \bar C (x)).
\end{array} \eqno(3.14)
$$
It is obvious now that the (anti-)co-BRST charges which are the generators
of
(anti-)co-BRST transformations in (2.8) are the {\it translation
generators} along
the Grassmannian directions of the six dimensional supermanifold. The
nilpotency
of these charges (ie $Q_{(a)d}^2 = 0$) {\it geometrically} corresponds to a
couple of
successive ($ (\partial/\partial \theta)^2
= (\partial/\partial\bar\theta)^2 = 0$)
translations along the Grassmannian directions of the six dimensional
supermanifold (cf section 7).\\

\noindent
{\bf 4 On-shell nilpotent (co-)BRST symmetries: chiral superfield
formalism}\\

\noindent
To provide the geometrical origin and interpretation for the
on-shell nilpotent (co-)BRST
symmetries ($s_{(d)b}$) and corresponding generators
($Q_{(d)b}$), we resort to the chiral superfield
formulation on the four $(4 + 2)$-dimensional supermanifold. To this end
in mind, first of all we generalize the generic local field
$\Psi (x) = (A_\mu (x), C (x), \bar C (x))$ of the Lagrangian density (2.1)
to a chiral ($\partial_{\theta} \tilde A_{M}
(x,\theta,\bar\theta) = 0$) supervector superfield $\tilde A^{(c)}_{M}
(x,\bar\theta)
= (B^{(c)}_\mu (x,\bar\theta), \Phi^{(c)} (x,\bar \theta),
\bar \Phi^{(c)} (x,\bar\theta))$
with the following super expansions along the Grassmannian
$\bar\theta$-direction of the supermanifold
$$
\begin{array}{lcl}
B^{(c)}_{\mu} (x,  \bar \theta) &=& A_{\mu} (x)
+\; \bar \theta\; R_{\mu} (x) \nonumber\\
\Phi^{(c)} (x,  \bar \theta) &=& C (x)
+ i\;\bar \theta \;{\cal B} (x)
\nonumber\\
\bar \Phi^{(c)} (x,  \bar \theta) &=& \bar C (x)
+ i\; \bar \theta \;b_3 (x).
\end{array} \eqno(4.1)
$$
There are a few relevant points which we summon here: (i) it is
obvious that, in the limit $\bar \theta \rightarrow 0$, we get back the
generic field $\Psi (x)$ of the Lagrangian density (2.1). (ii) In general,
a superfield on the six $(4+2)$-dimensional supermanifold is parametrized
by
the superspace variables $Z^M = (x^\mu, \theta,\bar\theta)$ as discussed
earlier.
However, for our present discussions,
we have chosen $Z^M_{(c)} = (x^\mu, \bar\theta)$ as the chiral limit of the
general $Z^M$. (iii) The specific choices in the expansion of the
superfields
$\Phi^{(c)} (x, \bar \theta)$ and $\bar \Phi^{(c)} (x, \bar\theta)$ have
been
guided by the transformations in (2.5) and (2.6).
(iv) The total number of degrees of freedom for
the odd fields $(R_\mu, C, \bar C)$ and even fields
$(A_\mu, b_3, {\cal B} = - i \partial_i b^{(1)}_i/\nabla^2)$
match in the above expansion for the sake of consistency with the basic
tenets of supersymmetry. (v) The auxiliary fields $R_\mu, b_3, {\bf
b^{(1)}}$ will be
fixed in terms of the basic fields after the application of the
(dual-)horizontality
conditions. Some of them can also be fixed by resorting to the
equations of motion for the Lagrangian density (2.4). (vi) All the
component
fields, on the r.h.s. of the above expansion, are functions of the
spacetime
even variable $x^\mu$ alone. (vii) The super expansions in (4.1) are the
chiral limit ($\theta \rightarrow 0$) of the most general expansions in
(3.1).
(viii) The auxiliary field ${\bf b^{(2)}}$ has not been taken into the
expansion
because its equivalent (the magnetic field ${\bf B}$) does not transform
under (anti-)BRST as well as (anti-)co-BRST transformations.

Now we exploit the horizontality condition
($\tilde F = (\tilde d \tilde A)|_{(c)} = d A = F$) w.r.t. (super) exterior
derivatives $(\tilde d) d$ in the construction of the (super) curvature
two-form. Physically, the requirement of the horizontality condition
implies an imposition that the curvature two-form in the ordinary
spacetime manifold remains unchanged and it is restricted not to get any
contribution from the {\it superspace} variables.
The explicit expressions for the l.h.s. and r.h.s. in the horizontality
condition
$(\tilde d \tilde A)|_{(c)} = d A$ are
$$
\begin{array}{lcl}
(\tilde d \tilde A)|_{(c)} &=& (dx^\mu \wedge dx^\nu) (\partial_\mu
B^{(c)}_\nu)
+ (dx^\mu \wedge d \bar\theta) ( \partial_\mu \Phi^{(c)} - \partial_{\bar
\theta}
B^{(c)}_\mu ) - (d \bar\theta \wedge d \bar \theta) (\partial_{\bar\theta}
\Phi^{(c)})
\nonumber\\
d A &=& (dx^\mu \wedge dx^\nu)\; (\partial_\mu A_\nu)
\equiv \frac{1}{2} (dx^\mu \wedge dx^\nu) (\partial_\mu A_\nu
- \partial_\nu A_\mu)
\end{array} \eqno(4.2)
$$
where the super exterior derivative (defined in terms of the chiral
superspace coordinates) and super connection one-form
(defined in terms of the chiral superfields) are
$$
\begin{array}{lcl}
\tilde d|_{(c)} &=& d Z^M_{(c)} \;\partial_{M} \equiv
d x^\mu \;\partial_\mu + d \bar \theta\;\partial_{\bar\theta}
\nonumber\\
\tilde  A|_{(c)} &=& d Z^M_{(c)}\; \tilde A^{(c)}_{M} \equiv
dx^\mu \;B^{(c)}_\mu (x,\bar\theta)
+ d \bar \theta\; \Phi^{(c)} (x,\bar\theta).
\end{array} \eqno(4.3)
$$
It is straightforward to check that the horizontality restriction
$(\tilde d \tilde A)|_{(c)} = d A$ leads to the following relationships
$$
\begin{array}{lcl}
\partial_{\bar\theta} \Phi^{(c)}
= 0 \rightarrow {\cal B} (x) \equiv - i {\displaystyle
\frac{\partial_i b^{(1)}_i (x)}{\nabla^2}} = 0 \;\qquad\;
\partial_{\bar\theta} B^{(c)}_\mu  = \partial_\mu \Phi^{(c)}
\rightarrow
R_\mu (x) = \partial_\mu C (x)
\end{array} \eqno(4.4)
$$
and the restriction $\partial_\mu B_\nu - \partial_\nu B_\mu =
\partial_\mu A_\nu - \partial_\nu A_\mu $ implies
$\partial_\mu R_\nu - \partial_\nu R_\mu = 0$ which is readily satisfied by
$R_\mu = \partial_\mu C$. It is obvious that the condition $(\tilde d
\tilde A)|_{(c)}
= d A$ does not fix the auxiliary field $b_3 (x)$ in terms of the basic
fields
of the Lagrangian density (2.1). However, the equation of motion for the
Lagrangian density (2.4) comes to our rescue as: $b_3 (x) = -
(\partial \cdot A) (x)$. With these substitutions for the auxiliary fields,
the super expansion (3.1) becomes:
$$
\begin{array}{lcl}
B_{\mu}^{(c)} (x,  \bar \theta) &=& A_{\mu} (x)
+\; \bar \theta\; \partial_{\mu} C (x)
\equiv A_\mu (x) + \bar\theta\; (s_{b} A_\mu (x)) \nonumber\\
\Phi^{(c)} (x,  \bar \theta) &=& C (x)
+ i\; \bar \theta \;({\cal B} (x) = 0)
\equiv C (x) + \bar\theta\; (s_{b} C (x) = 0) \nonumber\\
\bar \Phi^{(c)} (x,  \bar \theta) &=& \bar C (x)
- i \; \bar \theta \;(\partial \cdot  A) (x)
\equiv \bar C (x) + \bar\theta\; (s_{b} \bar C(x)).
\end{array} \eqno(4.5)
$$
In fact, now the on-shell nilpotent BRST symmetry transformations in (2.2)
can be concisely written in terms of the above superfields expansions as
$$
\begin{array}{lcl}
s_{b} B^{(c)}_\mu (x,\bar\theta) = \partial_\mu \Phi^{(c)} (x,\bar\theta)
\quad
s_{b} \Phi^{(c)} (x,\bar\theta) = 0 \quad
s_{b} \bar \Phi^{(c)} (x,\bar\theta) = - i\;
(\partial \cdot  B)^{(c)} (x,\bar\theta).
\end{array} \eqno(4.7)
$$
One can readily check that the first transformation in the above equation
leads to $s_{b} A_\mu = \partial_\mu C, s_{b} C = 0$; the second
transformation produces
$s_{b} C = 0$ and the third one generates $s_{b} \bar C = - i
(\partial \cdot  A), s_{b} (\partial \cdot A) = \Box C$ in terms of the
basic fields of Lagrangian density (2.1).
It is interesting to check, vis-a-vis equation (2.8),  that
$$
\begin{array}{lcl}
{\displaystyle \frac{\partial}{\partial \bar\theta}}\;
\tilde A^{(c)}_{M} (x,\bar\theta)
= - i \;\bigl [ \Psi (x), Q_{b} \bigr ]_{\pm} \equiv s_{b} \Psi (x) \quad
\tilde A^{(c)}_{M} = (\Phi, \bar \Phi, B_\mu)^{(c)} \quad \Psi = (C, \bar
C, A_\mu)
\end{array} \eqno(4.8)
$$
where the brackets $[\;,\;]_{\pm}$ stand for the (anti-)commutators when
the
generic field $\Psi$ and superfield $\tilde A^{(c)}_M$ are
(fermionic)bosonic
in nature. Thus, conserved and on-shell nilpotent BRST charge $Q_{b}$
geometrically turns out to be
the translation generator $\partial / \partial \bar \theta$
for the superfields $\tilde A^{(c)}_M$ along the $\bar\theta$-direction
of the supermanifold. The process of this translation generates
the on-shell nilpotent BRST symmetry transformations on $\Psi$
which correspond to (2.2). In addition, the nilpotency of $s_{b}^2 = 0$
and $Q_{b}^2 = 0$ is intimately connected with the property of the
square of the translational generator
(ie $(\partial/\partial\bar\theta)^2 = 0$).

We illustrate now the derivation of the on-shell nilpotent dual(co-)BRST
symmetry transformations of (2.3) by exploiting the analogue of the
horizontality condition
\footnote{ We christen this condition as the dual-horizontality condition
because $\tilde d (d)$ and $\tilde \delta (\delta)$ are dual
($\tilde \delta = - \star \tilde d \star, \delta = - * d *$)
to each-other. Here the Hodge duality operations $*$ and $\star$ are
defined on the 4D Minkowski manifold and 6D supermanifold, respectively.
The restriction $(\tilde \delta \tilde A)|_{(c)} = \delta A$ amounts
to setting equal to zero all the Grassmannian parts of the superscalar
$(\tilde \delta \tilde A)|_{(c)}$. On the ordinary even dimensional
manifold, the
operation $\delta A = - \;*\; d \;*\; A $ always yields the
covariant gauge-fixing term $(\partial \cdot A)$ (ie zero-form) for the
1-form ($A = dx^\mu A_\mu$) Abelian $U(1)$ gauge theory in any arbitrary
spacetime dimension.} w.r.t. (super) co-exterior derivatives
$(\tilde\delta) \delta$ by requiring $(\tilde \delta \tilde A)|_{(c)}
= \delta A$.
It is pretty obvious that the chiral limit (ie $\theta \rightarrow 0$)
of the most general expression for $\tilde \delta \tilde A$ in the
equation (3.9) yields the following expression for $(\tilde \delta \tilde
A)|_{(c)}$
$$
\begin{array}{lcl}
\mbox{Lim}_{\theta \rightarrow 0}\; (\tilde \delta \tilde A) =
(\tilde \delta \tilde A)|_{(c)}\equiv (\partial \cdot B)^{(c)}
(x,\bar\theta)
- s^{\bar\theta\bar\theta}
\partial_{\bar\theta}\; \bar \Phi^{(c)} (x,\bar\theta).
\end{array} \eqno(4.9)
$$
Due to the dual-horizontality requirement, the above equation
(defined on the supermanifold) is to be equated with
$\delta A = - * d * A \equiv (\partial \cdot A)$ (defined on the ordinary
4D spacetime
manifold). This restriction leads to the following relationships
$$
\begin{array}{lcl}
\partial_{\bar\theta} \bar \Phi^{(c)}
(x,\bar\theta) = 0 \rightarrow b_3 (x) = 0 \;\qquad\;
(\partial \cdot B)^{(c)} (x,\bar\theta) = (\partial \cdot A) (x)
\rightarrow
(\partial \cdot R) (x) = 0.
\end{array} \eqno(4.10)
$$
The above condition $(\partial \cdot R) = 0$ is satisfied automatically by
the choice
$R_0 = i \bar C, R_{i} = i (\partial_{i}\partial_0)/\nabla^2) \bar C$.
It is obvious that, in the expansion (4.1),  the auxiliary field 
${\cal B}  = - i
(\partial_{i} b^{(1)}_i/ \nabla^2) $ or
${\cal B}^{(I)}  = - i
(\partial_{i} b^{(1)}_i + e J_{0})/ \nabla^2)$ for the free as well as 
interacting theory
is not fixed in terms of the basic
fields of
(2.1) by the dual-horizontality condition. However,
the equation
of motion for the Lagrangian density (2.4) helps us to get ${\bf b^{(1)} =
E}$. Thus,
the chiral super expansion (3.1), for the {\it free } theory, becomes
$$
\begin{array}{lcl}
B^{(c)}_{0} (x,  \bar \theta) &=& A_{0} (x)
+ \bar \theta\; (i\;\bar C (x))
\equiv A_0 (x) + \bar\theta\; (s_{d} A_0 (x)) \nonumber\\
B^{(c)}_{i} (x,  \bar \theta) &=& A_{i} (x)
+ \bar \theta\;
\Bigl (
i {\displaystyle \frac{\partial_i \partial_0} {\nabla^2}}
\;\bar C (x)\Bigr )
\equiv A_i (x) + \bar\theta\; (s_{d} A_i (x)) \nonumber\\
\Phi^{(c)} (x,  \bar \theta) &=& C (x)
+ \bar \theta \;\Bigl ({\displaystyle  \frac{\partial_i E_i (x)}
{\nabla^2}} \Bigr )
\equiv C (x) + \bar\theta\; (s_{d} C (x)) \nonumber\\
\bar \Phi^{(c)} (x,  \bar \theta) &=& \bar C (x)
+  \bar \theta \;(i\;b_3 (x) = 0)
\equiv \bar C (x) + \bar\theta\; (s_{d} \bar C(x) = 0).
\end{array} \eqno(4.11)
$$
It is now evident that
$$
\begin{array}{lcl}
&&{\displaystyle \frac{\partial}{\partial \bar\theta}}\;
\tilde A^{(c)}_{M} (x,\bar\theta)
= - i \;\bigl [ \Psi (x), Q_{d} \bigr ]_{\pm} \equiv \;s_{d} \;\Psi (x)
\nonumber\\
&& \tilde A^{(c)}_{M} (x, \bar\theta) =
(\Phi, \bar \Phi, B_0, B_i)^{(c)} (x, \bar\theta)\qquad \Psi (x)
= (C , \bar C, A_0, A_i) (x)
\end{array} \eqno(4.12)
$$
where the brackets have the same meaning as discussed earlier.
This equation shows that {\it geometrically} the on-shell nilpotent co-BRST
charge $Q_{d}$ is the generator of translation $\partial/ \partial
\bar\theta$
for the chiral superfield $\tilde A^{(c)}_M$ along the Grassmannian
direction
$\bar\theta$ of the $(4+2)$-dimensional supermanifold. Furthermore, the
on-shell nilpotency conditions $s_{d}^2 = 0$ and $ Q_{d}^2 = 0$ are
connected
with the property of the square of the translational generator
$(\partial/\partial \bar\theta)^2 = 0$. The process of
the translation of $\tilde A^{(c)}_M (x,\bar\theta)
= (B_0, B_i, \Phi, \bar \Phi)^{(c)} (x,\bar\theta) $ along the
$\bar\theta$-direction
produces the on-shell nilpotent co-BRST transformation $s_{d} \Psi$ for the
generic local field $\Psi = (A_\mu, C, \bar C)$. Thus, there exists a
mapping,
namely; $s_{d} \leftrightarrow \partial/\partial\bar\theta$.

There is a clear-cut
distinction, however, between $Q_{b}$ and $Q_{d}$ as far as the translation
of
the fermionic superfields (or transformations on (anti-)ghost fields
$(\bar C)C$) along $\bar\theta$-direction is concerned. For instance,
the translation generated by $Q_{b}$ along $\bar\theta$-direction
results in the transformation for the anti-ghost field
$\bar C$ but analogous translation by $Q_{d}$ leads to the transformation
for the ghost field $C$. In more sophisticated language, the horizontality
condition entails upon the chiral superfield $\bar \Phi$ to remain chiral
but the chiral superfield $\Phi$ becomes a local spacetime field
(i.e., $\Phi (x,\bar\theta) = C (x)$). On the contrary, the
dual-horizontality
condition entails upon the chiral superfield $\Phi$ to retain its chirality
but
the chiral superfield $\bar\Phi$ becomes
an ordinary local field (ie $\bar\Phi (x,\bar\theta) = \bar C (x)$).\\

\noindent
{\bf 5 Anti-BRST and anti-co-BRST symmetries: anti-chiral superfields}\\

\noindent
To derive the on-shell nilpotent anti-BRST and anti-co-BRST symmetry
transformations of (2.2) and (2.3), we resort to the anti-chiral
superfields
$\tilde A^{(ac)}_M (x,\theta) = (B^{(ac)}_\mu, \Phi^{(ac)},
\bar \Phi^{(ac)}) (x,\theta)$ which have
the following super expansions along the $\theta$-direction of the
supermanifold
$$
\begin{array}{lcl}
B^{(ac)}_{\mu} (x,  \theta) &=& A_{\mu} (x)
+\; \theta\; \bar R_{\mu} (x) \nonumber\\
\Phi^{(ac)} (x,  \bar \theta) &=& C (x)
- i\; \theta \;b_3  (x) \nonumber\\
\bar \Phi^{(ac)} (x,  \theta) &=& \bar C (x)
+ i\; \theta \;\bar {\cal B} (x).
\end{array} \eqno(5.1)
$$
These are, in fact, the anti-chiral limit ($\bar\theta \rightarrow 0$)
of the general super expansion (3.1) on the
$(4+2)$-dimensional supermanifold with an exception. The change in sign of
the expansion for the superfield $\Phi^{(ac)} (x,\theta)$ has been taken
for the
algebraic convenience which amounts to the replacement $\bar b_3 (x)
\rightarrow - b_3 (x)$. This choice has been guided by our knowledge of
the most general discussion of nilpotent symmetries in section 2.
The super exterior derivative $\tilde d|_{(ac)}$
and super connection one-form $\tilde A|_{(ac)}$, for our present
discussion, are
$$
\begin{array}{lcl}
\tilde d|_{(ac)} &=& d Z^M_{(ac)} \;\partial_{M} \equiv
d x^\mu \;\partial_\mu + d  \theta\;\partial_{\theta}
\nonumber\\
\tilde  A|_{(ac)}  &=& d Z^M_{(ac)}\; \tilde A_{M} \equiv
dx^\mu \;B^{(ac)}_\mu (x,\theta)
+ d  \theta\; \bar \Phi^{(ac)} (x,\theta)
\end{array} \eqno(5.2)
$$
which are the anti-chiral limit ($\theta \rightarrow 0,
d \theta \rightarrow 0$) of the corresponding general expressions defined
in (3.2). Now the imposition of the horizontality condition $(\tilde d
\tilde A)|_{(ac)} = d A$ implies the restriction that the curvature
two-form
$F = d A$, defined on the ordinary spacetime manifold, remains unchanged.
In other words, the superspace contributions to the curvature
two-form are set equal to zero. The following inputs (ie the anti-chiral
limit of (3.3))
$$
\begin{array}{lcl}
(\tilde d \tilde A)|_{ac} &=& (dx^\mu \wedge dx^\nu) (\partial_\mu
B^{(ac)}_\nu)
+ (dx^\mu \wedge d \theta) ( \partial_\mu \bar \Phi^{(ac)} -
\partial_{\theta}
B^{(ac)}_\mu ) - (d \theta \wedge d \theta) (\partial_{\theta} \bar
\Phi^{(ac)})
\nonumber\\
d A &=& (dx^\mu \wedge dx^\nu)\; (\partial_\mu A_\nu) \equiv
\frac{1}{2}\; (dx^\mu \wedge dx^\nu)\; (\partial_\mu A_\nu - \partial_\nu
A_\mu)
\end{array} \eqno(5.3)
$$
lead to the following relationships due to $d A = (\tilde d \tilde
A)|_{ac}$
$$
\begin{array}{lcl}
\partial_{\theta} \bar \Phi^{(ac)} (x,\theta)
= 0 \rightarrow \bar {\cal B} (x) = 0 \qquad
\partial_\mu \bar \Phi^{(ac)} (x,\theta)
= \partial_\theta B^{(ac)}_\mu (x,\theta) \rightarrow \bar R_\mu (x)
= \partial_\mu \bar C (x)
\end{array} \eqno(5.4)
$$
and $\partial_\mu B^{(ac)}_\nu - \partial_\nu B^{(ac)}_\mu =
\partial_\mu A_\nu - \partial_\nu A_\mu $  which implies
$\partial_\mu \bar R_\nu - \partial_\nu \bar R_\mu = 0$. The latter
requirement
is automatically satisfied by $\bar R_\mu = \partial_\mu \bar C$. It is
clear
that the above horizontality restriction does not fix $b_3 (x)$ in terms of
the
basic fields of the Lagrangian density (2.1). However, the equation of
motion $ b_3 = - (\partial \cdot A)$ for the Lagrangian density (2.4) comes
to our help. With these insertions, the super expansion (5.1) becomes
$$
\begin{array}{lcl}
B^{(ac)}_{\mu} (x,  \theta) &=& A_{\mu} (x)
+\; \theta\; \partial_{\mu} \bar C (x)
\equiv A_\mu (x) + \theta\; (s_{ab} A_\mu (x)) \nonumber\\
\Phi^{(ac)} (x,  \bar \theta) &=& C (x)
+ i \; \theta \; (\partial \cdot A) (x)
\equiv C (x) + \theta\; (s_{ab} C (x)) \nonumber\\
\bar \Phi^{(ac)} (x,  \theta) &=& \bar C (x)
+ i\; \theta \;(\bar {\cal B} (x) = 0)
\equiv \bar C (x) + \theta \; (s_{ab} \bar C (x) = 0).
\end{array} \eqno(5.5)
$$
It is now straightforward to check that
$$
\begin{array}{lcl}
&&{\displaystyle \frac{\partial}{\partial \theta}}\; \tilde A^{(ac)}_{M}
(x,\theta)
= - i \;\bigl [ \Psi (x), Q_{ab} \bigr ]_{\pm}\; \equiv \;\;s_{ab} \Psi (x)
\nonumber\\
&&\tilde A^{(ac)}_{M} = (\Phi, \bar \Phi, B_\mu)^{(ac)}
\qquad\;\; \Psi = (C , \bar C, A_\mu)
\end{array} \eqno(5.6)
$$
where the above brackets have the same interpretation as discussed earlier.
This equation shows that {\it geometrically} the on-shell nilpotent
anti-BRST
charge $Q_{ab}$ is the generator of translation
$\partial / \partial \theta$
for the anti-chiral superfield $\tilde A^{(ac)}_{M} (x,\theta) =
(B_\mu, \Phi, \bar\Phi)^{(ac)} (x,\theta)$
along the $\theta$-direction of the supermanifold. In fact, this process of
translation induces the anti-BRST symmetry transformations
(ie $s_{ab} \Psi$) for the local fields $\Psi$ that are listed in equation
(2.2). Thus, there is a mapping $s_{ab} \leftrightarrow \partial /
\partial\theta$
between the above two key operators and the nilpotency
of the anti-BRST transformation $s_{ab}^2 = 0$ (as well as the
corresponding
charge $Q_{ab}^2 = 0)$) is encoded in the square of the translation
generator $(\partial/\partial\theta)^2 = 0$.

To dwell a bit on the derivation of the on-shell nilpotent
anti-co-BRST symmetry, we shall resort to
the anti-chiral superfield formulation. It is straightforward to check that
the anti-chiral limit ($ \theta \rightarrow 0$) of the most general
expression (3.9) leads to the following
$$
\begin{array}{lcl}
\mbox{Lim}_{\bar\theta \rightarrow 0}\;
(\tilde \delta \tilde A) = (\tilde \delta \tilde A)|_{(ac)}
\equiv (\partial \cdot B)^{(ac)}
- s^{\theta\theta} (\partial_{\theta} \Phi^{(ac)}).
\end{array} \eqno(5.7)
$$
The restriction $(\tilde \delta \tilde A)|_{(ac)} = \delta A$
(which physically implies an imposition that the zero-form gauge-fixing
term
$\delta A = (\partial \cdot A)$, defined on the ordinary
spacetime manifold, remains unchanged) leads to the following
relationships
$$
\begin{array}{lcl}
(\partial_{\theta}  \Phi^{(ac)}) (x,\theta) = 0 \rightarrow b_3 (x) = 0
\qquad
(\partial \cdot B)^{(ac)}
= (\partial \cdot A) \rightarrow (\partial \cdot \bar R) = 0.
\end{array} \eqno(5.8)
$$
The condition $(\partial \cdot \bar R) = 0$ is
readily satisfied by the choice $\bar R_0 = i C, R_i = i (\partial_0
\partial_i/
\nabla^2) C$. The dual-horizontality condition
$(\tilde \delta \tilde A)|_{(ac)} = \delta A$ does not fix the field
$\bar {\cal B}  = + i (\partial_i b^{(1)}_i / \nabla^2)$ or
$\bar {\cal B}^{(I)}  = + i (\partial_i b^{(1)}_i + e J_{0})/ \nabla^2$ 
in terms of the basic fields of free as well as interacting
theories. The equation of motion ${\bf b^{(1)} = E}$ for
the Lagrangian density (2.4), however, comes to our rescue. The super
expansion for the free 4D Abelian theory becomes
$$
\begin{array}{lcl}
B^{(ac)}_{0} (x,  \theta) &=& A_{0} (x)
+ \theta \; (i\; C (x))
\equiv A_0 (x) + \theta\; (s_{ad} A_0 (x)) \nonumber\\
B^{(ac)}_{i} (x,  \theta) &=& A_{i} (x)
+\; \theta \Bigl ({\displaystyle \frac{i \partial_0 \partial_i} {\nabla^2}}
\Bigr )\;C (x)
\equiv A_i (x) + \theta\; (s_{ad} A_i (x)) \nonumber\\
\Phi^{(ac)} (x,  \theta) &=& C (x)
+  \theta\; (i \;b_3 (x) = 0)
\equiv C (x) + \theta\; (s_{ad} C (x) = 0) \nonumber\\
\bar \Phi^{(ac)} (x,  \theta) &=& \bar C (x)
+ \theta \;\Bigl ({\displaystyle - \frac{ \partial_i E_i} {\nabla^2}}\Bigr
) (x)
\equiv \bar C (x) + \theta \; (s_{ad} \bar C (x)).
\end{array} \eqno(5.9)
$$
The geometrical interpretation for the co-BRST charge $Q_{ad}$ is encoded
in
$$
\begin{array}{lcl}
&&{\displaystyle \frac{\partial}{\partial \theta}}\; \tilde A^{(ac)}_{M}
(x,\theta)
=\; - i \;\bigl [ \Psi (x), Q_{ad} \bigr ]_{\pm} \;\equiv \;s_{ad} \Psi (x)
\nonumber\\
&& \tilde A^{(ac)}_{M} = (\Phi, \bar \Phi, B_0, B_i)^{(ac)}
\qquad  \;\Psi = (C , \bar C, A_0, A_i)
\end{array}\eqno(5.10)
$$
where the brackets $[\;,\;]_{\pm}$ have the same interpretation as
explained
earlier. It is obvious to note that $Q_{ad}$ turns out to be the
translation generator $(\partial /\partial\theta)$
for the anti-chiral superfields
$\tilde A^{(ac)}_M (x,\theta) = (B_\mu, \Phi, \bar \Phi)^{(ac)} (x,\theta)$
along the
$\theta$-direction of the supermanifold. The action of the
on-shell nilpotent transformation operator $s_{ad}$ on the local fields
$\Psi$ and the operation of $(\partial / \partial\theta)$
on the anti-chiral superfields
$\tilde A^{(ac)}_M (x,\theta)$ are inter-related and there exists a mapping
$s_{ad} \leftrightarrow (\partial / \partial\theta)$. The nilpotency
$s_{ad}^2= 0$ is connected to $(\partial/\partial\theta)^2 = 0$.
Even though both the charges
$Q_{ad}, Q_{ab}$ have the similar kind of mapping with the translation
generator, there is a clear distinction between them. Whereas the former
generates a transformation for the ghost field $C$ through the translation
of the superfield $\Phi$, the latter generates
the corresponding transformation on the anti-ghost field $\bar C$ through
the translation of $\bar \Phi$ superfield. The direction of
translation for the superfields is common for both of them
(ie the $\theta$-direction of the supermanifold).\\

\noindent
{\bf 6 On-shell nilpotent symmetries: general superfield formulation}\\

\noindent
It should be emphasized that the on-shell nilpotent (anti-)BRST-
and (anti-)co-BRST symmetries can be derived {\it together} if we
merge systematically the (anti-)chiral superfields and have the super
expansion for the {\it free} theory as (see, eg, [19] for details)
$$
\begin{array}{lcl}
B_{\mu} (x, \theta, \bar \theta) &=& A_{\mu} (x)
+ \theta\; \bar R_{\mu} (x) + \bar \theta\; R_{\mu} (x)
+ i \;\theta \;\bar \theta S_{\mu} (x) \nonumber\\
\Phi (x, \theta, \bar \theta) &=& C (x)
+ i\; \theta  (\partial \cdot  A) (x)
+ i \;\bar \theta\; \Bigl ({\displaystyle \frac{- i \partial_i E_i}
{\nabla^2}}
\Bigr ) (x)
+ i\; \theta\; \bar \theta \;s (x) \nonumber\\
\bar \Phi (x, \theta, \bar \theta) &=& \bar C (x)
+ i \;\theta\; \Bigl ({\displaystyle \frac{i \partial_i E_i}
{\nabla^2}}\Bigr ) (x)
+ i\; \bar \theta \;\bigl (-(\partial \cdot A) \bigr ) (x)
+ i \;\theta \;\bar \theta \;\bar s (x).
\end{array} \eqno(6.1)
$$
For the interacting theory, one has to replace $\partial_i E_i$ in the above
by $(\partial_i E_i + e J_{0})$.
In our earlier works [19,24], similar super expansions with the definitions
in (3.2) and $\star$ operation defined in (3.7) and (3.10)
(together with the ones given in the Appendix), have been exploited in the
horizontality condition ($\tilde F = \tilde d \tilde A = d A = F$)
as well as in the dual-horizontality conditions ($ \tilde \delta \tilde A
= \delta A)$ for the 2D free Abelian and self-interacting
non-Abelian gauge theories. For our present free as well as interacting
4D theory,
the horizontality condition $(\tilde d \tilde A = d A)$ leads
to the derivation of the auxiliary fields in terms of the basic fields of
the Lagrangian density (2.1) as follows
$$
\begin{array}{lcl}
R_\mu = \partial_\mu C \qquad \bar R_\mu = \partial_\mu \bar C

\qquad S_\mu = - \partial_\mu (\partial \cdot A) \qquad s = \bar s = 0.
\end{array}\eqno(6.2)
$$
Taking the help of the above expressions, the expansions in (6.1) can be
expressed in terms of the on-shell nilpotent (anti-)BRST symmetries (2.2)
as
$$
\begin{array}{lcl}
B_{\mu} (x, \theta, \bar \theta) &=& A_{\mu} (x)
+ \theta\; (s_{ab} A_\mu (x)) + \bar \theta\; (s_{b} A_\mu (x))
+  \;\theta \;\bar \theta (s_{b} s_{ab} A_\mu (x)) \nonumber\\
\Phi (x, \theta, \bar \theta) &=& C (x)
+ \; \theta  (s_{ab} C (x))
+ \;\bar \theta\; (s_{b} C (x))
+ \; \theta\; \bar \theta \;(s_{b} s_{ab} C (x)) \nonumber\\
\bar \Phi (x, \theta, \bar \theta) &=& \bar C (x)
+ \;\theta\; (s_{ab} \bar C (x)) + \; \bar \theta \;(s_{b} \bar C (x))
+  \;\theta \;\bar \theta \;(s_{b} s_{ab} \bar C(x)).
\end{array} \eqno(6.3)
$$
In a similar fashion, the dual horizontality condition ($\tilde \delta
\tilde A
= \delta A$) w.r.t. (super) co-exterior derivatives $(\tilde \delta)\delta$
leads to the following relationships
$$
\begin{array}{lcl}
(\partial \cdot R) = (\partial \cdot \bar R) = (\partial \cdot S)
= 0 \qquad s = \bar s = 0.
\end{array}\eqno(6.4)
$$
It is evident now that the above relations have solutions in (3.12) and
(3.13)
which satisfy all the conditions. Thus, in terms of the on-shell nilpotent
(anti-)co-BRST symmetry transformations (2.3), the expansion in (6.1) can
be
written as
$$
\begin{array}{lcl}
B_{\mu} (x, \theta, \bar \theta) &=& A_{\mu} (x)
+ \theta\; (s_{ad} A_\mu (x)) + \bar \theta\; (s_{d} A_\mu (x))
+  \;\theta \;\bar \theta (s_{d} s_{ad} A_\mu (x)) \nonumber\\
\Phi (x, \theta, \bar \theta) &=& C (x)
+ \; \theta  (s_{ad} C (x))
+ \;\bar \theta\; (s_{d} C (x))
+ \; \theta\; \bar \theta \;(s_{d} s_{ad} C (x)) \nonumber\\
\bar \Phi (x, \theta, \bar \theta) &=& \bar C (x)
+ \;\theta\; (s_{ad} \bar C (x)) + \; \bar \theta \;(s_{d} \bar C (x))
+  \;\theta \;\bar \theta \;(s_{d} s_{ad} \bar C(x)).
\end{array} \eqno(6.5)
$$
We would like to lay stress on the fact that  it is only for the free
(one-form) Abelian gauge theory that (anti-)chiral superfields are merged
together
systematically to produce the on-shell nilpotent (anti-)BRST and
(anti-)co-BRST symmetries together in the framework of the geometrical
superfield formulation. The same is not true for the non-Abelian
gauge theory in any arbitrary dimension of spacetime. In fact, the on-shell
nilpotent
anti-BRST and anti-co-BRST symmetries do not exist for the non-Abelian
gauge theories.
For the derivation of the off-shell nilpotent versions
of the (anti-)BRST and (anti-)co-BRST symmetries for the non-Abelian
gauge theories, one has to introduce
another set of auxiliary fields (see, eg, [18,25-28] for the details).\\

\noindent
{\bf  7 Conclusions}\\

\noindent
In the present investigation, we have derived the off-shell as well as
on-shell
nilpotent versions of the (anti-)BRST and (anti-)co-BRST symmetry
transformations  for the
free 4D one-form Abelian gauge theory in the framework of geometrical
superfield formalism. For this purpose, we have invoked general superfields
as well as (anti-)chiral superfields and their corresponding
super expansions. We have also derived a mapping between the
translation generators $(\partial/\partial\theta,
\partial/\partial\bar\theta)$
(along the $(\theta, \bar\theta)$ directions of the six ($4 +
2)$-dimensional
supermanifold) and the internal nilpotent transformations of the on-shell
variety
$s_{(a)b}, s_{(a)d}$ as well as the off-shell variety $\tilde s_{(a)b},
\tilde s_{(a)d}$
for the Lagrangian density of the theory, as
$$
\begin{array}{lcl}
&&{\displaystyle \frac{\partial}{\partial \bar\theta}} \leftrightarrow
s_{(d)b}
\;\;\;\;\;\qquad\;\;\;\;\;
{\displaystyle \frac{\partial}{\partial \theta}} \leftrightarrow s_{ab}
\;\;\;\;\;\qquad\;\;\;\;\;
{\displaystyle \frac{\partial}{\partial \theta}} \leftrightarrow s_{ad}
\nonumber\\
&&\tilde s_{(d)b} \leftrightarrow
\mbox{ Lim}_{\theta, \bar\theta \rightarrow 0}\;
{\displaystyle \frac{\partial}{\partial \bar \theta}}\; \qquad
\tilde s_{ab} \leftrightarrow
\mbox{ Lim}_{\theta, \bar\theta \rightarrow 0}\;
{\displaystyle \frac{\partial}{\partial \theta}}\; \qquad
\tilde s_{ad} \leftrightarrow
\mbox{ Lim}_{\theta, \bar\theta \rightarrow 0}\;
{\displaystyle \frac{\partial}{\partial \theta}}.
\end{array}\eqno(7.1)
$$
This mapping enables us to provide the {\it geometrical} interpretation for
the
conserved and nilpotent (anti-)BRST ($Q_{(a)b}$)
and (anti-)co-BRST $(Q_{(a)d})$ charges as the translation
generators $(\partial/\partial\theta,\partial/\partial\bar\theta)$
along the Grassmannian directions of the supermanifold. Furthermore, it
also
provides the {\it geometrical} origin and interpretation for the nilpotency
($Q_{(a)b}^2 = 0, Q_{(a)d}^2 = 0$) property of these charges as
a couple of successive translations (ie, ($\partial/\partial\theta)^2
= (\partial/\partial\bar\theta)^2 = 0$) along the Grassmannian directions
of the
supermanifold. The above statements are valid for the off-shell as well as
on-shell
versions of the (anti-)BRST ($Q_{(a)b}$)
and (anti-)co-BRST ($Q_{(a)d}$) charges and their specific
nilpotent properties.

One of the interesting features of our investigation is the observation
(and its proof) that the (dual-)horizontality conditions on the
(anti-)chiral
superfields lead to the derivation of the on-shell nilpotent (anti-)BRST
and (anti-)co-BRST symmetries (cf sections 4 and 5)
that co-exist together (cf section 6) for the Lagrangian density of a
4D free Abelian gauge theory. We have shown that (anti-)chiral superfields
can merge
consistently in the case of the free 4D Abelian gauge theories and they
lead to the derivation of the on-shell nilpotent (anti-)BRST and
(anti-)co-BRST
symmetries {\it together}. The same does not happen in the case of
self-interacting 2D non-Abelian gauge theory (see, eg, [24] for details).
As emphasized in the introduction, one of the key
differences between the (anti-)BRST and (anti-)co-BRST transformations
is the fact that whereas the former transformations are local, covariant,
continuous
and nilpotent, the latter are non-local, non-covariant, continuous and
nilpotent.
To capture the non-locality and non-covariance of the
latter transformations in the framework of superfield approach, we have
chosen the
non-local auxiliary fields ${\cal B} = - i (\partial_{i}
b^{(1)}_i/\nabla^2)$ and
$\bar {\cal B} = + i (\partial_{i} b^{(1)}_i/\nabla^2)$ in the super
expansion of
the superfields $\Phi (x,\theta,\bar\theta)$ and $\bar
\Phi(x,\theta,\bar\theta)$ for the free 4D Abelian gauge theory. For the
case of the interacting theory, one can choose instead:
${\cal B}^{(I)} = - i (\partial_{i}
b^{(1)}_i + e J_{0})/\nabla^2$ and
$\bar {\cal B}^{(I)} = + i (\partial_{i} b^{(1)}_i + e J_{0})/\nabla^2$.
In this context, it is worthwhile to mention an interesting observation in
[31]
that these non-locality and non-covariance disappear in the momentum phase
space
if we take into account the key ingredients and
inputs from the  Wigner's little group. In fact, the choice
of the reference frame $ k^\mu = (\omega, 0, 0,\omega)$ for the propagating
massless ($k^2 = 0)$ photon along the $z$-direction of the 4D manifold with
energy
$\omega$ simplifies all the (anti-)commutators of the theory and the whole
discussion
on the BRST cohomology becomes local and covariant in this particular
reference frame
(see, eg, [31] for details).

In the framework of superfield formalism, the non-locality and
non-covariance of
the transformations on the gauge field $A_\mu$ turns up from the conditions
$ (\partial \cdot R) = (\partial \cdot \bar R) = 0$ which emerge due to the
dual-horizontality condition (cf (3.9) and (3.11)). This is not the case
for
the two $(1 + 1)$-dimensional (2D) (i) free Abelian gauge theory [32-34],
(ii)
interacting Abelian gauge theory where $U(1)$ gauge field couples with the
Dirac fields [35,36], (iii) self-interacting non-Abelian gauge theory
[37,34], etc,
where the local and covariant solutions for the the conditions $(\partial
\cdot R)
= (\partial \cdot \bar R) = 0$ do exist as: $R_\mu = - \varepsilon_{\mu\nu}
\partial^\nu \bar C$ and $R_\mu = - \varepsilon_{\mu\nu} \partial^\nu C$
where
$\varepsilon_{\mu\nu}$ is the anti-symmetric Levi-Civita tensor in 2D with
$ \varepsilon_{01} = + 1 = \varepsilon^{10}$. Unlike the precise and
unique derivation of the
(anti-)BRST symmetry transformations due to the horizontality condition
$\tilde d \tilde A = d A$, the dual-horizontality condition $\tilde \delta
\tilde A
= \delta A$ does not exactly and uniquely lead to the derivation of all the
auxiliary
fields $R_\mu (x)$ and $\bar R_\mu (x)$ in terms of the (anti-)ghost fields
$(\bar C)C$. In fact, for the 4D theory,
one has to make judicious choice for $R_{0}, \bar R_{0},
R_{i}$ and $\bar R_i$ in terms of the anticommuting (anti-)ghost fields for
the validity
of the conditions $ (\partial \cdot R) = (\partial \cdot \bar R) = 0$. In
a similar fashion, one has to make judicious and clever guess for the
expression for $S_\mu$ (cf (3.13)) so that it can (i) satisfy $(\partial
\cdot S)
= 0$, and (ii) be consistent with expansions in (3.5) and (3.14). It can be
checked
that our choice in (3.13) fulfills both these criteria. In fact, the
non-uniqueness
of the solutions for $(\partial \cdot R) = (\partial \cdot \bar R) = 0$ in
the case
of 4D 1-form Abelian gauge theory is very interesting because it is
primarily
this feature of the superfield formulation which is
responsible for the existence of several guises of the
dual-BRST symmetries that has been discussed extensively in [4].
These different looking symmetries correspond
to different choices of $R's$ (and $\bar R's$)
such as $ R_{0} = i \nabla^2 \bar C, i (\nabla^2/ \partial_{0})
\bar C$ and $R_i = i \partial_0 \partial_i \bar C, i \partial_i \bar C$
(and $ \bar R_{0} = i \nabla^2 C, i (\nabla^2/\partial_{0}) C,
\bar R_i = i \partial_0 \partial_i C, i \partial_i C$) etc under which
the gauge-fixing term remains invariant. Thus, in some sense, the
superfield
formulation with the super co-exterior derivative $\tilde \delta$ and
the corresponding dual-horizontality condition do
provide the reason for the existence
of several forms of the
(non-local, non-covariant, continuous and nilpotent)
dual-BRST symmetries for the 1-form Abelian gauge theory.

It has been a long-standing problem to obtain, in a systematic way, the
BRST-type transformations (cf (2.2), (2.3), (2.5) (2.6)) on the matter
fields $\psi$ and $\bar \psi$ in the framework of superfield approach
applied to the BRST formalism. So far, the BRST-type transformations on the
gauge fields (see, eg, [12-17] for the 1-form and 2-form free gauge theories)
and the ghost fields have been obtained in the superfield formulation. In fact,
this is the present status of this approach because, hitherto, only the
(dual-)horizontality conditions $\tilde \delta \tilde A = \tilde \delta A,
\tilde d \tilde A = d A$ etc (that involve the (super-)gauge fields
and super (co-)exterior derivatives) have been exploited in the derivation
of the BRST-type symmetries on the gauge- and the ghost fields. In these
restrictions, it is obvious that the matter fields 
$\psi$ and $\bar\psi$ play no significant role at all.
This is the central reason that, so far, it has not been possible to obtain
the BRST-type symmetries on the matter fields in the superfield
approach. However, we strongly feel that, the continuity equation 
$\delta J = 0 \rightarrow \partial_\mu J^\mu = 0$, which involves 
the 1-form $J$
(ie $J = dx^\mu J_\mu$, with $J_\mu = \bar \psi \gamma_\mu \psi$) 
and the co-exterior
derivative $\delta$ (ie $\delta = - * d *$), might play an important role in
the derivation of the BRST-type transformations on the matter fields.
In this restriction, all one has to do is to have the super expansions
for the superfields corresponding to the matter fields $\psi$ and $\bar \psi$
analogous to (3.1). The insertions of these superfields in
the restriction $(\tilde \delta \tilde J = \delta J = 0$), corresponding
to the continuity equation $(\partial_\mu J^\mu = 0$), might lead to
the derivation of BRST-type transformations on the matter fields. 
There is another clue which might help in such a derivation. This is
connected with the restriction that the interaction term
$J^\mu A_\mu$ should remain unchanged in the process of supersymmetrization. 
In other words, this amounts to the condition : $\tilde J^\mu B_\mu
= J^\mu A_\mu$ where $\tilde J^\mu$ is the current constructed with
the superfields corresponding to the matter fields $\psi$ and $\bar \psi$
and $B_\mu$ is the superfield defined in (3.1).
These issues and ideas are  under investigation at the  
moment and the preliminary results are found to be encouraging for QED in 2D.

It is interesting to point out that local, covariant, continuous and
(off-shell as well as on-shell) nilpotent versions of the (anti-)BRST and
(anti-)co-BRST
symmetries have been obtained for the 4D free Abelian 2-form gauge theory
defined
on the flat Minkowski manifold [38,39]. Its quasi-topological nature has
been
discussed in [39] and it has been shown that this theory provides a
tractable
field theoretical model for the Hodge theory in 4D [38,39]. The
``extended''
BRST cohomology for this theory has been discussed in [40] where the
insights from
the Wigner's little group play a very crucial role. It would be interesting
endeavour
to capture the (anti-)BRST and (anti-)co-BRST symmetries for the above
2-form Abelian
gauge theory in the framework of superfield formalism and provide
geometrical
origin for the nilpotent charges in the theory. Such studies might shed
some
light on the quasi-topological nature (see, eg, [39]) of this theory in the
framework of superfield formalism and it might provide some clue to attempt
the nilpotent symmetries of this kind present in the case of non-Abelian
2-form gauge
theories. All such understandings of the 2-form gauge theories will furnish
some
insights into our main goal of understanding the interacting 2-form gauge
theories
where there is a gauge invariant
coupling between the matter fields and the antisymmetric
(2-form) gauge field. Another very interesting endeavour that can be
attempted is based on the local $OSp(4|2)$ supersymmetry and its connection
with
the extended BRST transformations in the context of gravitational theories
where the geometrical
superfield approach could be applied (see, eg, [15] for details).
In fact, the extension of our work to the realm of gravitational theories
will complete the generality of the application of super co-exterior
derivative and the corresponding dual-horizontality
condition. These are some of the issues which  are under investigation
and our results will be reported elsewhere [41].\\

\noindent
{\bf Acknowledgements}\\

\noindent
Fruitful and enlightening discussions with L. Bonora (SISSA),
K. S. Narain (AS-ICTP) and G. Thompson (AS-ICTP)
are gratefully acknowledged. Thanks are also due
to members of HEP group at AS-ICTP for their warm hospitality
and stimulating conversations. It is a pleasure to acknowledge
some clarifying and constructive suggestions by the adjudicator.\\

\begin{center}

{\bf Appendix}\\

\end{center}

\noindent
In addition to the $\star$ operations in (3.7) and (3.10), we collect
here some of the $\star$ operations on the wedge products of the super
differentials
defined on the six $(4 + 2)$-dimensional supermanifold. We have followed
our convention
of putting the spacetime differentials to the left and the Grassmannian
differentials
to the right in every wedge products. Some of these $\star$ operations are
$$
\begin{array}{lcl}
&&\star\; (dx^\mu \wedge dx^\nu\wedge dx^\lambda\wedge d\theta)
= \varepsilon^{\mu\nu\lambda\xi}\;\; (dx_\xi \wedge d \bar\theta)
\nonumber\\
&&\star\; (dx^\mu \wedge dx^\nu\wedge dx^\lambda\wedge d \bar \theta)
= \varepsilon^{\mu\nu\lambda\xi}\;\; (dx_\xi \wedge d\theta) \nonumber\\
&&\star\; (dx_\mu\wedge d\bar\theta)
= \frac{1}{3!}\;\varepsilon_{\mu\nu\lambda\xi}\;\;
(dx^\nu\wedge dx^\lambda\wedge dx^\xi \wedge d \theta)\nonumber\\
&&\star\; (dx_\mu\wedge d\theta)
= \frac{1}{3!}\;\varepsilon_{\mu\nu\lambda\xi}\;\;
(dx^\nu\wedge dx^\lambda\wedge dx^\xi \wedge d \bar\theta)\nonumber\\
&&\star\; (dx^\mu \wedge dx^\nu\wedge d\theta\wedge d\theta)
= \frac{1}{2!}\;\varepsilon^{\mu\nu\lambda\xi}\; \;
(dx_\lambda \wedge dx_\xi) s^{\theta\theta}\nonumber\\
&&\star\;[(dx_\mu\wedge dx_\nu) s^{\theta\theta}]
= \frac{1}{2!}\;\varepsilon_{\mu\nu\lambda\xi}\;\;
(dx^\lambda\wedge dx^\xi\wedge d\theta\wedge d\theta)\nonumber\\
&&\star\; (dx^\mu \wedge dx^\nu\wedge d\bar\theta\wedge d\bar\theta)
= \frac{1}{2!}\;\varepsilon^{\mu\nu\lambda\xi}\; \;
(dx_\lambda \wedge dx_\xi) \;s^{\bar\theta\bar\theta}\nonumber\\
&&\star\;[(dx_\mu\wedge dx_\nu)\; s^{\bar\theta\bar\theta}]
= \frac{1}{2!}\;\varepsilon_{\mu\nu\lambda\xi}\;\;
(dx^\lambda\wedge dx^\xi\wedge d\bar\theta\wedge d\bar\theta)\nonumber\\
&&\star\; (dx^\mu \wedge d\theta\wedge d\theta)
= \frac{1}{3!}\;\varepsilon^{\mu\nu\lambda\xi}\;\;
(dx_\nu \wedge dx_\lambda \wedge dx_\xi) \;s^{\theta\theta}\nonumber\\
&&\star\;[(dx_\mu\wedge dx_\nu\wedge dx_\lambda)\; s^{\theta\theta}]
= \varepsilon_{\mu\nu\lambda\xi}\;\;
(dx^\xi\wedge d\theta\wedge d\theta)\nonumber\\
&&\star\; (dx^\mu\wedge d\bar\theta\wedge d\bar\theta)
= \frac{1}{3!}\;\varepsilon^{\mu\nu\lambda\xi}\;\;
(dx_\nu\wedge dx_\lambda \wedge dx_\xi)\;
s^{\bar\theta\bar\theta}\nonumber\\
&&\star\;[(dx_\mu\wedge dx_\nu \wedge dx_\lambda)
\;s^{\bar\theta\bar\theta}]
= \varepsilon_{\mu\nu\lambda\xi}\;\;
(dx^\xi\wedge d\bar\theta\wedge d\bar\theta) \nonumber\\
&&\star\; (d\theta\wedge d\theta)
= \frac{1}{4!}\;\varepsilon^{\mu\nu\lambda\xi}\;\;
(dx_\mu \wedge dx_\nu\wedge dx_\lambda \wedge dx_\xi)\;
s^{\theta\theta}\nonumber\\
&&\star\;[(dx_\mu\wedge dx_\nu \wedge dx_\lambda \wedge dx_\xi)
\;s^{\theta\theta}]
= \varepsilon_{\mu\nu\lambda\xi}\;\;
( d\theta\wedge d\theta) \nonumber\\
&&\star\; (d\bar\theta\wedge d\bar\theta)
= \frac{1}{4!}\;\varepsilon^{\mu\nu\lambda\xi}\;\;
(dx_\mu \wedge dx_\nu\wedge dx_\lambda \wedge dx_\xi)\;
s^{\bar\theta\bar\theta}\nonumber\\
&&\star\;[(dx_\mu\wedge dx_\nu \wedge dx_\lambda \wedge dx_\xi)
\;s^{\bar\theta\bar\theta}]
= \varepsilon_{\mu\nu\lambda\xi}\;\;
(d\bar\theta\wedge d\bar\theta).
\end{array}
$$
It should be noted that we have not included some of the $\star$ operations
on
the super differentials containing ($d\theta \wedge d\bar\theta$) because,
as
pointed out in section 3, these can arise in two entirely different ways.
While taking
the $\star$ of such differentials, one has to be careful about the presence
and/or absence of $s^{\theta\bar\theta}$ as illustrated in (3.10). We would
like
to emphasize that we have chosen here some of the super differentials where
some
non-trivialities are present. However, one can exploit easily the above
understanding to take the
$\star$ operations on differentials of the form $(dx^\mu \wedge dx^\nu)$
etc
where only spacetime differentials are present. These operations
would be analogous to what we have already done in equation (3.7).

\end{document}